\documentclass[10pt,twoside,twocolumn]{IEEEtran}%
\usepackage{latexsym}
\usepackage{amsmath}
\usepackage{graphicx}
\usepackage{epsfig}%
\usepackage{amsfonts}%
\usepackage{amssymb}
\newtheorem{theorem}{Theorem}
\newtheorem{lemma}[theorem]{Lemma}

\newtheorem{corollary}[theorem]{Corollary}

\begin{document}

\title{Recursive Decoding and Its Performance for Low-Rate Reed-Muller Codes}
\date{}
\author{Ilya Dumer\thanks{This work was supported by the National Science
Foundation under Grant CCR-0097125.  The material in this paper
was presented in part at the 37th Allerton Conf. on Communication,
Control, and Computing, Monticello, IL, USA, Sept. 22-24, 1999.
\newline The author is with the College of Engineering, University of
California, Riverside, CA 92521, USA (e-mail: dumer@ee.ucr.edu).
}} \maketitle

\begin{abstract}
Recursive decoding techniques are considered for Reed-Muller (RM)
codes of growing length $n$ and fixed order $r.$ An algorithm is
designed that has complexity of order $n\log n$ and corrects most
error patterns of weight up to $n(1/2-\varepsilon)$ given that
$\varepsilon$ exceeds $n^{-1/2^{r}}.$ This improves the asymptotic
bounds known for decoding RM codes with nonexponential complexity.

To evaluate decoding capability, we develop a probabilistic
technique that disintegrates decoding into a sequence of recursive
steps. Although dependent, subsequent outputs can be tightly
evaluated under the assumption that all preceding decodings are
correct. In turn, this allows us to employ the second-order
analysis and find the error weights for which the decoding error
probability vanishes on the entire sequence of decoding steps as
the code length $n$ grows.


\end{abstract}

\section{Introduction}

In this paper, our goal is to design new decoding algorithms that
can enhance techniques known to date for RM codes. \ In general,
RM codes can be designed from the set $\left\{
f_{\,\,r}^{m}\right\}  $ of all $m$-variate Boolean polynomials of
degree $r$ or less. Here each polynomial $f$ $\in$\ $\left\{
f_{\,\,r}^{m}\right\}  $ is defined on the $m$-dimensional space
$E_{2}^{m}.$ For any $f,$ we consider the sequence of binary
values \ $f(x)$ obtained as argument $x$ runs through $E_{2}^{m}$.
These sequences - codewords $\mathbf{c}(f)$ - form an RM code,
which is below denoted $\left\{
\genfrac{}{}{0pt}{}{m}{r}%
\right\}  $ and has length $n,$ dimension $k,$ and distance $d$ as follows:%
\[
n=2^{m},\quad k=\sum_{i=0}^{r}(_{i}^{m}),\quad d=2^{m-r}.
\]
The decoding algorithms discussed in this paper (including the new algorithms)
can be applied to any RM code. However, we will mostly focus on their
asymptotic performance obtained for long RM codes of \textit{fixed order} $r.$
To define their error-correcting performance, we use the following definition.
Given an infinite sequence of codes $A_{i}(n_{i},d_{i}),$ we say that a
decoding algorithm $\Psi$ has a \textsf{threshold }sequence $\delta_{i}$ and a
\textsf{residual} sequence $\epsilon_{i}\rightarrow0$ if for $n_{i}%
\rightarrow\infty:\smallskip$

$\bullet$ $\Psi$ correctly decodes all but a vanishing fraction of
error patterns of weight $\delta_{i}(1-\epsilon_{i})$ or less;

$\bullet$ $\Psi$ fails to decode a nonvanishing fraction of error patterns of
weight $\delta_{i}$ or less\footnote{Note that multiple sequences with
$\epsilon_{i}\rightarrow0$ can satisfy the same definition$.$}$.\smallskip$

Nonexponential decoding algorithms known for RM codes can be loosely separated
into three groups. First, \textsf{majority decoding}\textit{\ }was developed
in the seminal paper \cite{ree}. The algorithm requires complexity of order
$nk$ or less. \ For RM codes of fixed order $r,$ it was proven in \cite{kri}
that majority decoding achieves maximum possible threshold $\delta=n/2$ (here
and below we omit index $i)$ with a residual
\begin{equation}
\varepsilon_{r}^{\text{maj}}=(cm/d)^{1/2^{r+1}},\quad m\rightarrow\infty,
\label{eps-maj}%
\end{equation}
where $c$ is a constant that does not depend on $m$ and $r.$

The second type of decoding algorithms makes use of the symmetry group of RM
codes. One very efficient algorithm is presented in \cite{sid}. For long RM
codes $\left\{
\genfrac{}{}{0pt}{}{m}{2}%
\right\}  $ this algorithm reduces the residual term $\varepsilon
_{2}^{\text{maj}}$ from (\ref{eps-maj}) to its square $(cm/d)^{1/4},$ where
$c>\ln4$. On the other hand, the complexity order of $nm^{2}$ of majority
decoding is also increased in algorithm \cite{sid} to almost its square
$n^{2}m.$ The corresponding thresholds for higher orders $r\geq3$ are yet unknown.

Another result of \cite{sid} concerns maximum-likelihood (ML) decoding. It is
shown that ML decoding of RM codes of fixed order $r$ yields a substantially
lower residual
\begin{equation}
\varepsilon_{r}^{\text{min}}=m^{r/2}n^{-1/2}(c(2^{r}-1)/r!)^{1/2},\quad
m\rightarrow\infty, \label{ml}%
\end{equation}
where $c>\ln4.$ However, even the best known algorithm of ML decoding designed
by the multilevel trellis structure in \cite{for} has yet complexity that is
exponent in $n$.

Finally, various \textsf{recursive} techniques were introduced in \cite{lit},
\cite{hemm}, \cite{kab}, and \cite{bos}. All these algorithms use different
recalculation rules but rely on the same code design based on
the\textit{\ Plotkin construction} $(\mathbf{u,u+v}).$ The construction allows
to decompose RM codes $\left\{
\genfrac{}{}{0pt}{}{m}{r}%
\right\}  $ onto shorter codes, by taking subblocks\textbf{\ }$\mathbf{u}$
\ and $\mathbf{v}$ from codes $\left\{
\genfrac{}{}{0pt}{}{m-1}{r}%
\right\}  $ and $\left\{
\genfrac{}{}{0pt}{}{m-1}{r-1}%
\right\}  .$ The results from \cite{lit}, \cite{kab}, and \cite{bos} show that
this recursive structure enables both encoding and bounded distance decoding
with the lowest complexity order of $n\min(r,m-r)$ known for RM codes of an
arbitrary order $r$.

In the same vein, below we also employ Plotkin construction. The basic
recursive procedure will split RM code $\left\{
\genfrac{}{}{0pt}{}{m}{r}%
\right\}  $ of length $n$ into two RM codes of length $n/2$. Decoding is then
relegated further to the shorter codes of length $n/4$ and so on, until we
reach basic codes of order $r\leq1$ or $r=m.$ At these points, we use maximum
likelihood decoding or the variants derived therefrom. By contrast, in all
intermediate steps, we shall only recalculate the newly defined symbols. Here
our goal is to find efficient \textit{recalculation rules} that can
\textit{provably} improve the performance of RM codes. Our results presented
below in Theorems \ref{th:1-2} and \ref{th:1-1} show that recursive techniques
indeed outperform other polynomial algorithms known for RM codes. These
results also show how decoding complexity can be traded for a higher
threshold.${\smallskip}$

\begin{theorem}
\label{th:1-2} Long RM codes $\left\{
\genfrac{}{}{0pt}{}{m}{r}%
\right\}  $ of fixed order $r$ can be decoded with linear complexity $O(n)$
and decoding threshold%
\[
\delta=n/2,\quad\varepsilon_{r}=((2r\ln m)/d)^{1/2^{r+1}},\quad m\rightarrow
\infty.
\]
${\smallskip}$
\end{theorem}

\begin{theorem}
\label{th:1-1} Long RM codes $\left\{
\genfrac{}{}{0pt}{}{m}{r}%
\right\}  $ of fixed order $r$ can be decoded with quasi-linear complexity
\ $O(n\log n)$ and decoding threshold%
\[
\delta=n/2,\quad\tilde{\varepsilon}_{r}=(cm/d)^{1/2^{r}},\quad c>\ln4,\quad
m\rightarrow\infty.
\]
${\smallskip}$
\end{theorem}

\noindent Rephrasing Theorems \ref{th:1-2} and \ref{th:1-1}, we obtain the
following${\smallskip\smallskip}$

\begin{corollary}
Long RM codes $\left\{
\genfrac{}{}{0pt}{}{m}{r}%
\right\}  $ of fixed order $r$ can be decoded with vanishing
output error probability and linear complexity $O(n)$\ (or
quasi-linear complexity $O(n\log n))$ on a binary channel with
crossover error probability $(1-\varepsilon_{r})/2$ \
(correspondingly, $(1-\tilde{\varepsilon}_{r})/2)$ as
$n\rightarrow \infty.{\smallskip}$
\end{corollary}

Note that Theorem \ref{th:1-2} increases decoding threshold of the recursive
techniques introduced in \cite{lit} and \cite{kab} from the order of $d/2$ to
$n/2$ while keeping linear decoding complexity. Theorem \ref{th:1-1} improves
both the complexity and residual of majority decoding of RM codes. When
compared with the algorithm of \cite{sid}, this theorem reduces the quadratic
complexity $O(n^{2}\log n)$ to a quasi-linear complexity $O(n\log n)$ and also
extends this algorithm to an arbitrary order $r\geq2$ of RM codes.

The algorithms designed below differ from the former algorithms of
\cite{lit}, \cite{kab}, and \cite{bos} in both the intermediate
recalculations and the stopping rules. Firstly, we employ \ new
\textit{intermediate recalculations}, which yield the exact
decoding thresholds, as opposed to the bounded distance threshold
$d/2$ established in \cite{kab} and \cite{bos}. This leads us to
Theorem \ref{th:1-2}. Secondly, by analyzing the results of
Theorem \ref{th:1-2}, we also change the former \textit{stopping
rules}, all of which terminate decoding at the repetition codes.
Now we terminate decoding earlier, once we achieve the
biorthogonal codes. This change yields Theorem \ref{th:1-1} and
substantially improves decoding performance (this is discussed in
Section 7). Finally, we employ a new probabilistic analysis of
recursive algorithms. In Section 7, we will see that this analysis
not only gives the actual thresholds but also shows how the
algorithms can be advanced further.

Below in Section 2 we consider recursive structure of RM codes in
more detail. In Section 3, we proceed with decoding techniques and
design two different recursive algorithms $\Psi_{\,r}^{m}$ and
$\Phi_{\,r}^{m}.$ These algorithms are analyzed in Sections 4, 5,
and 6, which are concluded with Theorems \ref{th:1-2} and
\ref{th:1-1}. In Section 7, we briefly discuss extensions that
include decoding lists, subcodes of RM codes, and soft decision
channels. For the latter case, we will relate the noise power to
the quantity $\varepsilon^{-2}.$ Thus, the residual $\varepsilon$
will serve as a measure of the highest noise power that can be
withstood by a specific low-rate code.

\section{Recursive encoding of RM codes}

Consider any $m$-variate Boolean polynomial $f=f_{\,\,r}^{m}$ and the
corresponding codeword $\mathbf{c}(f)$\ with symbols $f(x).$ Below we assume
that positions $x=(x_{1},...,x_{m})$ are ordered lexicographically, with
$x_{1}$ being the senior digit. Note that any polynomial $f$ can be split as
\begin{equation}
f_{\,\,r}^{m}(x_{1},...,x_{m})=f_{\,\,\,\,\,\,r}^{m-1}(x_{2},...,x_{m}%
)+x_{1}f_{\,\,r-1}^{m-1}(x_{2},...,x_{m}), \label{poly}%
\end{equation}
where we use the new polynomials $f_{\,\,\,\,\,\,r}^{m-1}$ and $f_{\,\,r-1}%
^{m-1}.$ These polynomials are defined over $m-1$ variables and have degrees
at most $r$ and $r-1,$ respectively. Correspondingly, one can consider two
codewords $\mathbf{u=c(}f_{\,\,\,\,\,\,r}^{m-1})$\ and $\mathbf{v=c(}%
f_{\,\,r-1}^{m-1})$ that belong to the codes $\left\{
\genfrac{}{}{0pt}{}{m-1}{r}%
\right\}  $ and $\left\{
\genfrac{}{}{0pt}{}{m-1}{r-1}%
\right\}  $. Then representation (\ref{poly}) converts any codeword
$\mathbf{c}(f)\in$ $\left\{
\genfrac{}{}{0pt}{}{m}{r}%
\right\}  $ to the form $(\mathbf{u,u+v}).$ This is the well known
\textsf{Plotkin construction. }\

By continuing this process on codes $\left\{
\genfrac{}{}{0pt}{}{m-1}{r}%
\right\}  $ and $\left\{
\genfrac{}{}{0pt}{}{m-1}{r-1}%
\right\}  ,$ we obtain RM codes of length $2^{m-2}$ and so on. Finally, we
arrive at the end nodes, which are repetition codes $\left\{
\genfrac{}{}{0pt}{}{g}{0}%
\right\}  $ for any $g=1,...,m-r$ \ and full spaces $\left\{
\genfrac{}{}{0pt}{}{h}{h}%
\right\}  $ for any $h=1,...,r.$ This is schematically shown in Fig. 1 for RM
codes of length 8.

Now let $\mathbf{a}_{\,r}^{m}=\{a_{j}|j=1,k\}$ be a block of $k\ $information
bits $a_{j}$ \ that encode a vector $(\mathbf{u},\mathbf{u}+\mathbf{v}).$ By
decomposing this vector into $\mathbf{u}$ and $\mathbf{v,}$ we also split
$\mathbf{a}_{\,r}^{m}$ into two information subblocks $\mathbf{a}%
_{\,\,\,\,\,\,r}^{m-1}$ and $\mathbf{a}_{\,\,r-1}^{m-1}$ that encode vectors
$\mathbf{u}$ and $\mathbf{v,}$ respectively. In the following steps,
information subblocks are split further, until we arrive at the end nodes
$\left\{
\genfrac{}{}{0pt}{}{g}{0}%
\right\}  $ or $\left\{
\genfrac{}{}{0pt}{}{h}{h}%
\right\}  $. This is shown in Fig. 2. Note that only one information bit is
assigned to the left-end (repetition) code $\left\{
\genfrac{}{}{0pt}{}{g}{0}%
\right\}  ,$ while the right-end code $\left\{
\genfrac{}{}{0pt}{}{h}{h}%
\right\}  $ includes $2^{h}$ bits. Below these $2^{h}$ bits will be encoded
using the unit generator matrix. Summarizing, we see that any codeword can be
encoded from the information strings assigned to the end nodes $\left\{
\genfrac{}{}{0pt}{}{g}{0}%
\right\}  $ or $\left\{
\genfrac{}{}{0pt}{}{h}{h}%
\right\}  $, by repeatedly combining codewords $\mathbf{u} $ and $\mathbf{v}$
in the $(\mathbf{u},\mathbf{u}+\mathbf{v})$-construction.\smallskip

\ \ \ \ \ \ \ \ \ \ \ \ \ \ \ \ \ \ \ \ \ \ \ \ \ \ \ \ \ \ \ \ \ \ {\small 0,0}%
\ \ \ \ \ \ \ \ \ \ \ \ \ \ \ \ \ \ \ \ \ \ \ \ \ \ \

\ \ \ \ \ \ \ \ \ \ \ \ \ \ \ \ \ \ \ \ \ \ \ \ \ \ \ \ \ \ \ $\nearrow
\ \nwarrow$\ \ \ \ \ \ \ \ \ \ \ \ \ \ \ \ \ \ \ \ \ \ \ \ \ \

\ \ \ \ \ \ \ \ \ \ \ \ \ \ \ \ \ \ \ \ \ \ \ \ \ \ \ {\small 1,0}%
{\large \ \ \ }\ \ \ {\large \ }\ {\small 1,1 }\ \ \ \ \ \ \ \ \ \ \ \ \ \ \ \ \ \ \ \ \ \ \

\ \ \ $\ \ \ \ \ \ \ \ \ \ \ \ \ \ \ \ \ \ \ \ \ \nearrow$ \ $\nwarrow
${\large \ }$\ ${\large \ \ }\ $\nearrow$ $\ \nwarrow$ $\ \ \ \ \ \ \ \ \ \ \ \ \ \ \ \ \ $

\ \ \ \ \ \ \ \ \ \ \ \ \ \ \ \ \ \ \ \ {\small 2,0}{\large \ \ \ }%
\ {\large \ \ \ \ }\ {\small 2,1}{\large \ \ \ }\ {\large \ \ \ }\ {\small 2,2 \ \ \ \ \ \ \ \ \ \ \ \ \ \ \ \ \ \ \ \ \ \ }

\ \ \ \ \ \ \ \ \ \ \ \ \ \ \ \ \ $\nearrow$ $\ \nwarrow${\large \ \ \ \ }%
\ $\nearrow$ \ $\ \nwarrow${\large \ \ \ }\ $\nearrow$ $\ \nwarrow$\ \ \ \ \ \ \ \ \

\ \ \ \ \ \ \ \ \ \ \ \ \ {\small 3,0}{\large \ \ \ \ \ \ \ \ }{\small 3,1}%
{\large \ \ \ \ \ \ \ \ }\ {\small 3,2}{\large \ \ \ }\ {\large \ \ \ }%
\ {\small 3,3 \ \ \ \ \ \ \ \ \ \ \ }\vspace{0.1in}

\quad Fig. 1$:$ Decomposition of RM codes of length 8.\vspace{0.1in}

\noindent Given any algorithm $\psi,$ in the sequel we use notation $|\psi|$
\ for its complexity. Let $\psi_{\,r}^{m}$ denote the encoding \ described
above for the code $\left\{
\genfrac{}{}{0pt}{}{m}{r}%
\right\}  .$ Taking a complexity estimate from \cite{lit} and its enhancement
from \cite{kab}, we arrive at the following lemma.\smallskip

\ \ \ \ \ \ \ \ \ \ \ \ \ \ \ \ \ \ \ \ \ \ \ \ \ \ \ \ \ \ \ \ \ $\mathbf{a}%
_{\,r}^{m}$\ \ \ \ \ \ \ \ \ \ \ \ \ \

\ \ \ \ \ \ \ \ \ \ \ \ \ \ \ \ \ \ \ \ \ \ \ \ \ \ \ \ \ $\swarrow
$\ $\ \ \ \ \ \searrow$\ \ \ \ \ \ \ \ \ \ \

\ \ \ \ \ \ \ \ \ \ \ \ \ \ \ \ \ \ \ \ \ {\large \ \ }$\mathbf{a}%
_{\,\,r-1}^{m-1}${\large \ }\ \ \ \ \ \ \ \ $\ \mathbf{a}_{\,\,\,\,\,\,r}%
^{m-1}$\ \ \ \ \ \ \ \ \ \

$\ \ \ \ \ \ \ \ \ \ \ \ \ \ \ \ \ \ \ \ \ \swarrow\searrow$%
{\large \ \ \ \ \ \ \ \ \ \ \ \ \ }\ $\swarrow\searrow$ $\ \ \ \ \ \ \ \ \ $

\ \ \ \ \ \ \ \ \ \ \ \ \ \ \ $\mathbf{a}_{\,\,r-2}^{m-2}${\large \ \ \ }%
\ $\mathbf{a}_{\,\,r-1}^{m-2}${\large \ \ \ \ }$\mathbf{a}_{\,\,r-1}^{m-2}%
${\large \ \ \ \ }$\mathbf{a}_{\,\,\,\,\,r}^{m-2}$

\ \ \ \ \ \ \ \ \ \ \ \ \ \ $\swarrow\searrow${\large \ \ \ \ \ \ }%
$\swarrow\searrow$ \ \ \ {\large \ }$\swarrow\searrow$\ {\large \ \ \ \ }%
$\swarrow\searrow${\large \ }\vspace{0.1in}

\qquad Fig. 2. Decomposition of information paths\smallskip

\begin{lemma}
RM codes $\left\{
\genfrac{}{}{0pt}{}{m}{r}%
\right\}  $ can be recursively encoded with complexity
\begin{equation}
\left|  \psi_{\,r}^{m}\right|  \leq n\min(r,m-r). \label{encoding}%
\end{equation}
\end{lemma}

\noindent\textit{Proof.} First, note that the end nodes $\left\{
\genfrac{}{}{0pt}{}{g}{0}%
\right\}  $ and $\left\{
\genfrac{}{}{0pt}{}{h}{h}%
\right\}  $ require no encoding and therefore satisfy the complexity bound
(\ref{encoding}). Second, we verify that code $\left\{
\genfrac{}{}{0pt}{}{m}{r}%
\right\}  $ satisfies (\ref{encoding}) if the two constituent codes do.\ Let
the codewords $\mathbf{u}\in\left\{
\genfrac{}{}{0pt}{}{m-1}{r}%
\right\}  $ and $\mathbf{v}\in\left\{
\genfrac{}{}{0pt}{}{m-1}{r-1}%
\right\}  $ have encoding complexity $\left|  \psi_{\,\,\,\,\,\,r}%
^{m-1}\right|  $ and $\left|  \psi_{\,\,r-1}^{m-1}\right|  $ that satisfies
(\ref{encoding}). Then their $(\mathbf{u,u+v})$-combination requires
complexity
\[
\left|  \psi_{\,\,r}^{m}\right|  \leq\left|  \psi_{\,\,r-1}^{m-1}\right|
+\left|  \psi_{\,\,\,\,\,\,r}^{m-1}\right|  +n/2,
\]
where $n/2$ extra additions ($\operatorname{mod}2)$ were included to find the
right half $\mathbf{u+v}$. Now we substitute estimates (\ref{encoding}) for
quantities $\left|  \psi_{\,\,r-1}^{m-1}\right|  $ and $\left|  \psi
_{\,\,\,\,\,\,r}^{m-1}\right|  $. If $r<m-r,$ then
\[
|\psi_{\,r}^{m}|\leq n(r-1)/2+nr/2+n/2=nr.
\]
\ The two other cases, namely $r>m-r$ and $r=m-r,$ can be treated
similarly. \hfill$\square\smallskip$

Now consider an information bit $a_{j}$ associated with a left node $\left\{
\genfrac{}{}{0pt}{}{g}{0}%
\right\}  ,$ where $g\in\lbrack1,m-r].$ We will map $a_{j}$ onto a specific
``binary path''%
\[
\xi\overset{\text{def}}{=}(\xi_{1},...,\xi_{m})
\]
of length $m$ leading from the origin $\left\{
\genfrac{}{}{0pt}{}{m}{r}%
\right\}  $ to the end node \ $\left\{
\genfrac{}{}{0pt}{}{0}{0}%
\right\}  $. To do so, we first define the senior bit
\[
\xi_{1}=\left\{
\begin{array}
[c]{ll}%
0, & \text{if}\;a_{j}\in\mathbf{a}_{\,\,r-1}^{m-1},\smallskip\smallskip\\
1, & \text{if}\;a_{j}\in\mathbf{a}_{\,\,\,\,\,\,r}^{m-1}.
\end{array}
\right.
\]
Next, we take $\xi_{2}=0$ if $a_{j}$ encodes the left descendant subcode on
the following step. Otherwise, $\xi_{2}=1.$ Similar procedures are then
repeated at the steps $t=3,...,m-g,$ and give some \textsf{end subpath }
$\underline{\xi}$ that arrives at the node $\left\{
\genfrac{}{}{0pt}{}{g}{0}%
\right\}  .$ We then add $g$ right-hand steps and obtain a full path $\xi$
\ of length $m$ that arrives at the node $\left\{
\genfrac{}{}{0pt}{}{0}{0}%
\right\}  .$ Using notation $1^{g}$ for the sequence of $g$ ones,
we write
\[
\xi=(\,\underline{\xi}\,,1^{g}).
\]

Now consider any right-end node $\left\{
\genfrac{}{}{0pt}{}{h}{h}%
\right\}  $, where $h\in\lbrack1,r]$ and let $\underline{\xi}$ be
any right-end path that ends at this node. Then $\underline{\xi}$
is associated with $2^{h}$ \ information bits. Therefore we extend
$\underline{\xi}$ to the full length $m$ by adding any binary
suffix $(\xi_{m-h+1},...,\xi_{m}).$ This allows us to consider
separately all $2^{h}$ information bits and use common notation
$a(\xi)$.

When all left- and right-end paths are considered together, we
obtain all paths of length $m$ and binary weight $m-r$ or more.
This gives one-to-one mapping between $k$ information bits and
extended paths $\xi.$ Below all $\xi$ are ordered
lexicographically, as $m$-digital binary numbers.

\section{\ Recursive decoding}

Now we turn to recursive decoding algorithms. \ We map any binary
symbol $a$ onto $(-1)^{a}$ and assume that all code vectors belong
to $\{1,-1\}^{n}.$ Obviously, the sum $a+b$ of two binary symbols
is being mapped onto the product of their images. Then we consider
any codeword
\[
\mathbf{c=}(\mathbf{u,uv})
\]
transmitted over a binary symmetric channel with crossover probability
$p<1/2.$ The received block $\mathbf{y\in}\{1,-1\}^{n}$ consists of two halves
$\mathbf{y}^{\prime}$ and $\mathbf{y}^{\prime\prime}$, which are the corrupted
images of vectors $\mathbf{u}$ and $\mathbf{uv}$. We start with a basic
algorithm $\Psi_{\text{rec}}(\mathbf{y)}$ that will be later used in recursive
decoding. In our decoding, vector $\mathbf{y}$ will be replaced by the vectors
whose components take on real values from the interval $[-1,+1].$ Therefore we
take a more general approach and assume that $\mathbf{y\in}\mathbb{R}%
^{n}.\smallskip$

\textbf{Step 1}. \ We first try to find the codeword $\mathbf{v}$ from
$\left\{
\genfrac{}{}{0pt}{}{m-1}{r-1}%
\right\}  \mathbf{.}$ In the absence of noise, we have the equality
$\mathbf{v=y}^{\prime}\mathbf{y}^{\prime\prime}$ (which gives the binary sum
of vectors $\mathbf{y}^{\prime}$ and $\mathbf{y}^{\prime\prime}$ in the former
notation). On a noisy channel, we first find the ``channel estimate''%
\begin{equation}
\mathbf{y}^{v}=\mathbf{y}^{\prime}\mathbf{y}^{\prime\prime} \label{1}%
\end{equation}
of $\mathbf{v.}$ Next, we employ (any) decoding $\Psi(\mathbf{y}^{v}),$ which
will be specified later. The output is some vector $\hat{\mathbf{v}}%
\in\left\{
\genfrac{}{}{0pt}{}{m-1}{r-1}%
\right\}  $ and its information block
$\hat{\mathbf{a}}^{v}$.$\smallskip$

\textbf{Step 2}. \ We try to find the block\ $\mathbf{u\in}\left\{
\genfrac{}{}{0pt}{}{m-1}{r}%
\right\}  $ given $\hat{\mathbf{v}}$ from Step 1. Here we take two
corrupted versions of vector $\mathbf{u}$, namely
$\mathbf{y}^{\prime}$ in the left half and
$\mathbf{y}^{\prime\prime}\hat{\mathbf{v}}$ in the right half.
These two \textit{real }vectors are added and combined in their
``midpoint''
\begin{equation}
\mathbf{y}^{u}=(\mathbf{y}^{\prime}+\mathbf{y}^{\prime\prime}\hat
{\mathbf{v}})/2. \label{3}%
\end{equation}
Then we use some decoding $\Psi(\mathbf{y}^{u}),$ which is also
specified later. The output is some vector
$\hat{\mathbf{u}}\in\left\{
\genfrac{}{}{0pt}{}{m-1}{r}%
\right\}  $ and its information block $\hat{\mathbf{a}}^{u}.$ So,
decoding
$\Psi_{\text{rec}}(\mathbf{y)}$ is performed as follows.$\smallskip$%

\[
\frame{$%
\begin{array}
[c]{l}%
\text{Algorithm }\Psi_{\text{rec}}(\mathbf{y)}.\medskip\\
\text{1. Calculate vector }\mathbf{y}^{v}=\mathbf{y}^{\prime}\mathbf{y}%
^{\prime\prime}\text{.}\\
\text{Find }\mathbf{\hat{v}=}\Psi(\mathbf{y}^{v})\text{ and }\hat
{\mathbf{a}}^{v}.\medskip\\
\text{2. Calculate vector }\mathbf{y}^{u}=(\mathbf{y}^{\prime}+\mathbf{y}%
^{\prime\prime}\hat{\mathbf{v}})/2\text{.}\\
\text{Find }\mathbf{\hat{u}=}\Psi(\mathbf{y}^{u})\text{ and }\hat
{\mathbf{a}}^{u}.\medskip\\
\text{3. Output decoded components:}\\
\hat{\mathbf{a}}:=(\hat{\mathbf{a}}^{v}\mid\hat{\mathbf{a}}%
^{u});\quad\mathbf{\hat{c}}:=(\mathbf{\hat{u}}\mid\mathbf{\hat{u}\hat{v}}).
\end{array}
$}%
\]

\noindent In a more general scheme $\Psi_{\,r}^{m}$, we repeat this recursion
by decomposing subblocks $\mathbf{y}^{v}$ and $\mathbf{y}^{u}$ further. On
each intermediate step, we only recalculate the newly defined vectors
$\mathbf{y}^{v}$ and $\mathbf{y}^{u}$ \ using (\ref{1}) when decoder moves
left and (\ref{3}) when it goes right. Finally, vectors $\mathbf{y}^{v}$ and
$\mathbf{y}^{u}$ are decoded, once we reach the end nodes $\left\{
\genfrac{}{}{0pt}{}{g}{0}%
\right\}  $ and $\left\{
\genfrac{}{}{0pt}{}{h}{h}%
\right\}  $. Given any end code $C$ of length $l$ and any estimate
$\mathbf{z}\in\mathbb{R}^{l},$ we employ the (soft decision)
\textsf{minimum-distance} (MD) decoding $\Psi(\mathbf{z})=\hat{\mathbf{c}%
}$ that outputs a codeword $\hat{\mathbf{c}}$ closest to
$\mathbf{z}$ in the \textsf{Euclidean metric}. Equivalently,
$\hat{\mathbf{c}}$ maximizes
the inner product $\mathbf{(c,z).}$ The algorithm is described below.%

\[
\frame{$%
\begin{array}
[c]{l}%
\text{Algorithm }\Psi_{\,r}^{m}(\mathbf{y)}.\medskip\\
\text{1. If }0<r<m\text{, perform }\Psi_{\text{rec}}(\mathbf{y)}\text{ }\\
\text{using }\Psi(\mathbf{y}^{v})=\Psi_{\,\,r-1}^{m-1}\text{ and }%
\Psi(\mathbf{y}^{u})=\Psi_{\,\,\,\,\,r}^{m-1}\text{.}\medskip\\
\text{2. If }r=0,\text{ perform MD decoding }\\
\Psi(\mathbf{y}^{v})\text{ for code }\left\{
\genfrac{}{}{0pt}{}{r}{0}%
\right\}  .\medskip\\
\text{3. If }r=m,\text{ perform MD decoding }\\
\Psi(\mathbf{y}^{u})\text{ for code }\left\{
\genfrac{}{}{0pt}{}{r}{r}%
\right\}  .
\end{array}
$}%
\]

\noindent In the following algorithm $\Phi_{\,r}^{m}$, we refine algorithm
$\Psi_{\,r}^{m}(\mathbf{y)}$ by terminating decoding $\Psi$ at the
biorthogonal codes $\left\{
\genfrac{}{}{0pt}{}{g}{1}%
\right\}  $.%

\[
\frame{$%
\begin{array}
[c]{l}%
\text{Algorithm }\Phi_{\,r}^{m}(\mathbf{y)}.\medskip\\
\text{1. If }1<r<m\text{, perform }\Psi_{\text{rec}}(\mathbf{y)}\text{ }\\
\text{using }\Psi(\mathbf{y}^{v})=\Phi_{\,\,r-1}^{m-1}\text{ and }%
\Psi(\mathbf{y}^{u})=\Phi_{\,\,\,\,\,r}^{m-1}\text{.}\medskip\\
\text{2. If }r=1,\text{ perform MD decoding }\\
\Phi(\mathbf{y}^{v})\text{ for code }\left\{
\genfrac{}{}{0pt}{}{r}{1}%
\right\}  .\medskip\\
\text{3. If }r=m,\text{ perform MD decoding }\\
\Phi(\mathbf{y}^{u})\text{ for code }\left\{
\genfrac{}{}{0pt}{}{r}{r}%
\right\}  .
\end{array}
$}%
\]

\noindent Thus, procedures $\Psi_{\,r}^{m}$ and $\Phi_{\,r}^{m}$ have a
recursive structure that calls itself until MD decoding is applied on the end
nodes. Now the complexity estimate follows.$\smallskip$

\begin{lemma}
For any RM code $\left\{
\genfrac{}{}{0pt}{}{m}{r}%
\right\}  ,$ algorithms $\Psi_{\,r}^{m}$ and $\Phi_{\,r}^{m}$ have decoding
complexity%
\begin{align}
|\Psi_{\,r}^{m}|  &  \leq 4n\min(r,m-r)+n,\label{comp-fi}\\
|\Phi_{\,r}^{m}|  &  \leq 3n\min(r,m-r)+n(m-r)+n. \label{comp-f}%
\end{align}
\smallskip
\end{lemma}

\noindent\textit{Proof.} First, note that for trivial codes
$\left\{
\genfrac{}{}{0pt}{}{r}{0}%
\right\} $ and $\left\{
\genfrac{}{}{0pt}{}{r}{r}%
\right\},  $ MD decoding can be executed in $n$ operations and
satisfies the bound (\ref{comp-fi}) (here we assume that finding
the sign of a real value requires one operation). For biorthogonal
codes, their MD decoding $\Phi_{\,1}^{m}$ can be executed in
$n\log_{2}n+n+3$ operations using the Green machine or
$n\log_{2}n+2n$ operations using fast Hadamard transform (see
\cite{gre} or \cite{MS}, section 14.4). Obviously, this decoding
satisfies the upper bound (\ref{comp-f}).

Second, for both algorithms $\Psi$ and $\Phi$, vector
$\mathbf{y}^{v}$ in (\ref{1}) can be calculated in $n/2$
operations while vector $\mathbf{y}^{u}$ in (\ref{3}) requires
$3n/2$ operations. Therefore our decoding complexity satisfies the
same recursion
\begin{align*}
|\Psi_{\,r}^{m}|  &  \leq\left|  \Psi_{r-1}^{m-1}\right|  +\left|
\Psi_{\,\,\,\,\,r}^{m-1}\right|  +2n,\\
|\Phi_{\,r}^{m}|  &  \leq\left|  \Phi_{r-1}^{m-1}\right|  +\left|
\Phi_{\,\,\,\,r}^{m-1}\right|  +2n.
\end{align*}
Finally,  we verify that (\ref{comp-fi}) and (\ref{comp-f})
satisfy the above recursion, similarly to the derivation of \
(\ref{encoding}). \hfill $\square\smallskip$

\textit{Discussion.}

Both algorithms $\Psi_{\,r}^{m}$ and $\Phi_{r}^{m}$ admit bounded distance
decoding. This fact can be derived by adjusting the arguments of \cite{kab}
for our recalculation rules (\ref{1}) and (\ref{3}). Algorithm $\Psi_{\,r}%
^{m}$ is also similar to decoding algorithms of \cite{lit} and \cite{bos}.
However, our \textit{recalculation rules} are different from those used in the
above papers. For example, the algorithm of \cite{bos} performs the so-called
``min-sum'' recalculation%
\begin{equation}
\mathbf{y}^{v}=\text{sign}(\mathbf{y}^{\prime}\mathbf{y}^{\prime\prime}%
)\,\min\{\,|\mathbf{y}^{\prime}|\,,\,|\mathbf{y}^{\prime\prime}|\,\},
\label{boss}%
\end{equation}
instead of (\ref{1}). This (simpler) recalculation (\ref{1}) will allow us to
substantially expand the \textit{``provable''} decoding domain versus the
bounded-distance domain established in \cite{kab} and \cite{bos}. We then
further extend this domain in Theorem \ref{th:1-1}, also using the new
\textit{stopping rule} that replaces $r=0$ in $\Psi_{\,r}^{m}$ with $r=1$ in
$\Phi_{r}^{m}.$ However, it is yet an open problem to find the decoding domain
using any other recalculation rule, say those from \cite{lit}, \cite{kab},
\cite{bos}, or \cite{dum4}.

Finally, note that the scaling factor $1/2$ in recalculation rule (\ref{3}) brings
any component $y^{u}$ back to the interval $[-1,+1]$ used before this recalculation.
This scaling will also allow us to simplify some proofs, in particular that of
Lemma \ref{lm:nei1}. However, replacing (\ref{3}) by the simpler rule
$$\mathbf{y}^{u}=\mathbf{y}^{\prime}+\mathbf{y}^{\prime\prime}\hat{\mathbf{v}}$$
does not change any decoding results. Though
being equivalent, the new rule also reduces complexity estimates
(\ref{comp-fi}) and
(\ref{comp-f}) to%
\begin{align}
|\Psi_{\,r}^{m}|  &  \leq3n\min(r,m-r)+n,\label{com-fi}\\
|\Phi_{\,r}^{m}|  &  \leq2n\min(r,m-r)+n(m-r)+n. \label{com-f}%
\end{align}
These reductions in complexity notwithstanding, in the sequel we still use the original
recalculation rule (\ref{3}) for the only reason to simplify our proofs.

\section{Analysis of recursive algorithms}

\subsection{Intermediate outputs}

We begin with the algorithm $\Psi_{\,r}^{m}$ and later will use a similar
analysis for the algorithm $\Phi_{\,r}^{m}.$ Note that $\Psi_{\,r}^{m}$ enters
each end node multiple times, by taking all paths leading to this node. It
turns out that the output bit error rate (BER) significantly varies on
different nodes and even on different paths leading to the same node.
Therefore our first problem is to fix a path $\xi$ and estimate the output BER
for the corresponding information symbol $a(\xi).$ In particular, we will
define the most error-prone paths.

Consider any (sub)path of some length $s.$ Let $\underline{\xi}$
be its prefix of length $s-1,$ so that
$\xi=(\underline{\xi},\xi_{s}),$
where%
\begin{equation}
\xi=(\xi_{1},...,\xi_{s}),\quad\underline{\xi}=(\xi_{1},...,\xi_{s-1}),\quad
s\in\lbrack1,m]. \label{sub}%
\end{equation}
First, note that algorithm $\Psi_{\,r}^{m}(\mathbf{y)}$ repeatedly
recalculates its input $\mathbf{y}$, by taking either an estimate
$\mathbf{y}^{v}$ from (\ref{1}) when a path $\xi$ turns left or $\mathbf{y}%
^{u}$ \ from (\ref{3}) otherwise. The following lemma shows that recursive
decoding follows lexicographic order of our paths $\xi\in$ $\Gamma.$\smallskip

\begin{lemma}
For two paths $\gamma$ and $\xi,$ the bit $a(\xi)$ is decoded after
$a(\gamma)$ if \ $\xi>\gamma.$\smallskip
\end{lemma}

\noindent\textit{Proof.} \ Given two paths $\xi$ and $\gamma,$ let
$s$ be the first (senior) position where they disagree. If \
$\xi>$ $\gamma,$ then $\xi_{s}=1$ and $\gamma_{s}=0.$ Thus, after
$s$ steps, $\gamma$ moves left while $\xi$ moves right.
Correspondingly, $\gamma$ proceeds first.\hfill
$\square$\smallskip

On any subpath $\xi$ of length $s,$ algorithm $\Psi_{\,r}^{m}(\mathbf{y)}$
outputs some vector $\mathbf{y}(\xi)$ of length $2^{m-s}.$ Next, we derive a
recursive expression for $\mathbf{y}(\xi)$ using formulas (\ref{1}) and
(\ref{3}). In any step $s$, the algorithm first splits $\mathbf{y}%
(\,\underline{\xi}\,)$ into halves
$\mathbf{y}^{\prime}(\,\underline{\xi}\,)$ and
$\mathbf{y}^{\prime\prime}(\,\underline{\xi}\,).$ For $\xi_{s}=0,$
$\mathbf{y}(\xi)$ is given by recursion (\ref{1}) and is rewritten
below in the upper line of (\ref{main}).

If $\xi_{s}=1,$ then $\mathbf{y}(\xi)$ is obtained from (\ref{3}).
Here we also need the vector $\hat{\mathbf{v}}(\xi)$ decoded on
the preceding subpath $(\,\underline{\xi}\,,0)$. The corresponding
output is written in the second line of (\ref{main}):
\begin{equation}
\mathbf{y}(\xi)=\left\{
\begin{array}
[c]{ll}%
\mathbf{y}^{\prime}(\,\underline{\xi}\,)\mathbf{y}^{\prime\prime}%
(\,\underline{\xi}\,), & \text{if}\;\xi_{s}=0,\\
\mathbf{y}^{\prime}(\,\underline{\xi}\,)/2+\hat{\mathbf{v}}%
(\xi)\mathbf{y}^{\prime\prime}(\,\underline{\xi}\,)/2, & \text{if}\;\xi_{s}=1.
\end{array}
\right.  \label{main}%
\end{equation}
Finally, consider any left-end path $\xi=(\,\underline{\xi}\,,1^{g})$ that
passes some repetition code $\left\{
\genfrac{}{}{0pt}{}{g}{0}%
\right\}  $. Note that no preceding decodings are used after $\xi$ reaches the
repetition code $\left\{
\genfrac{}{}{0pt}{}{g}{0}%
\right\}  .$ Here we define the end result on the path $\xi$ as
\begin{equation}
y(\xi)\overset{\text{def}}{=}\sum_{i=1}^{2^{g}}y_{i}(\,\underline{\xi
}\,)/2^{g}, \label{md2}%
\end{equation}
by taking $\hat{\mathbf{v}}(\xi)=1$ in the last $g$ steps of
recursion (\ref{main}). Note that MD decoding also makes its
decision on the entire sum of symbols $y_{i}(\,\underline{\xi}\,)$
and outputs the symbol\footnote{Below we assume that sign(0) takes
values +1 and -1 with probability 1/2.}
\begin{equation}
\hat{a}(\xi)=\text{sign}(y(\xi)). \label{md1}%
\end{equation}
For any right-end code $\left\{
\genfrac{}{}{0pt}{}{h}{h}%
\right\}  ,$  the output is some vector $\mathbf{y}(\underline{\xi
})$ of length $2^{h}.$ Again, MD decoding takes every symbol
$y(\xi)$ on the full path $\xi$ and converts it into the
information bit $\hat{a}(\xi),$ making bit-by-bit decision
(\ref{md1})$.$ This is summarized as $\smallskip$

\begin{lemma}
For any end path $\xi,$ the algorithm $\Psi_{\,r}^{m}$ decodes the
outputs $y(\xi)$ into the information bits $\hat{a}(\xi)$ using
the rule (\ref{md1}).
\end{lemma}

\subsection{Conditional error probabilities}

Next, we consider the decoding error probability $P(\xi)$ for any
information bit $a(\xi).$ On an additive binary symmetric channel,
$P(\xi)$ does not depend on the transmitted codeword $\mathbf{c}$
and we can assume that $\mathbf{c=1}$. \ According to our decoding
rule (\ref{md1}), an error event $\{\hat{a}(\xi)=-1\}$ has
probability
\[
P(\xi)=\Pr\{y(\xi)<0\}.
\]
(here we follow footnote 2 and assume that $y(\xi)<0$ with probability 1/2 if
$y(\xi)=0.$)

Note, however, that recursive output $y(\xi)$ depends on the
outputs $\mathbf{v(}\gamma)$ obtained on all preceding paths
$\gamma<\xi.$ To simplify our calculations, we wish to consider
the above event $\{y(\xi)<0\}$ conditioned that all preceding
decodings are correct$.$ This implies that any path $\gamma<\xi$
gives an information bit $\hat{a}(\gamma)$ and a codeword
$\mathbf{v(}\gamma)$ as follows:
\[
\hat{a}(\gamma)=1,\quad\mathbf{v(}\gamma)=\mathbf{1.}%
\]
This assumption also allows us to simplify our recalculations (\ref{1}),
(\ref{3}), and (\ref{main}) by removing all vectors $\ \mathbf{v(}\gamma)$:
\begin{equation}
\mathbf{y}^{v}=\mathbf{y}^{\prime}\mathbf{y}^{\prime\prime},\quad
\mathbf{y}^{u}=(\mathbf{y}^{\prime}+\mathbf{y}^{\prime\prime})/2,\medskip
\label{main1}%
\end{equation}%
\begin{equation}
\mathbf{y}(\xi)=\left\{
\begin{array}
[c]{ll}%
\mathbf{y}^{\prime}(\,\underline{\xi}\,)\mathbf{y}^{\prime\prime}%
(\,\underline{\xi}\,), & \text{if}\;\xi_{s}=0,\\
\mathbf{y}^{\prime}(\,\underline{\xi}\,)/2+\mathbf{y}^{\prime\prime
}(\,\underline{\xi}\,)/2, & \text{if}\;\xi_{s}=1.
\end{array}
\right.  \smallskip\label{main2}%
\end{equation}
Therefore our first goal is to find how much unconditional probabilities
$P(\xi)$ change given that preceding decodings are correct. First, let
\begin{equation}
\xi_{\ast}=(0^{r},1^{m-r}) \label{w1}%
\end{equation}
be the leftmost path that begins with $r$ zeros followed by $m-r$ ones. For
any path $\xi$, let $A(\xi)$ and $B(\xi)$ denote the events
\begin{align}
A(\xi)  &  =\cap_{\gamma\leq\xi}\{\hat{a}(\gamma)=1\};\label{events}\\
B(\xi)  &  =\cap_{\gamma<\xi}\{\hat{a}(\gamma)=1\};\nonumber
\end{align}
which include all error vectors that are correctly decoded on the
paths $\gamma\leq\xi$ or $\gamma<\xi,$ respectively. We define the
complete ensemble of all error vectors  by $B(\xi_{\ast}).$  In
the sequel, we replace each probability $P(\xi)$ by the
probability
\begin{equation}
p(\xi)\overset{\text{def}}{=}\Pr\{y(\xi)<0|B(\xi)\}=\Pr\{\overline{A}%
(\xi)|B(\xi)\} \label{tot4}%
\end{equation}
conditioned that all previous decodings are correct. The following
upper bound (\ref{tot2}) conservatively assumes that an
information symbol $\hat {a}(\xi)$ is \textit{always} incorrect
whenever a failure occurs in any step $\gamma\leq\xi$. Similarly,
the upper bound in (\ref{tot3}) uses the formula of total
probability and adds up probabilities $p(\xi)$ over all paths
$\xi$. By contrast, the lower bound takes into account that the
block is always incorrect given the decoding failure on the first
step $\xi_{\ast}$.\smallskip

\begin{lemma}
\label{lm:prob0}For any path $\xi\in\Gamma,$ \ its bit error rate $P(\xi)$
\ satisfies inequality%
\begin{equation}
P(\xi)\leq\sum_{\gamma\leq\xi}p(\gamma). \label{tot2}%
\end{equation}
Block error probability $P$ \ satisfies inequalities
\begin{equation}
p(\xi_{\ast})\leq P\leq\sum_{\xi\in\Gamma}p(\xi). \label{tot3}%
\end{equation}
$\smallskip$
\end{lemma}

\noindent\textit{Proof.} \ The probability $P(\xi)$ can be estimated as%
\begin{align*}
P(\xi)  &  \leq\Pr\{\overline{A}(\xi)\}=\sum_{\gamma\leq\xi}\Pr\{\overline
{A}(\gamma)\cap B(\gamma)\}\\
&  \leq\sum_{\gamma\leq\xi}\Pr\{\overline{A}(\gamma)|B(\gamma)\}=\sum
_{\gamma\leq\xi}p(\gamma).
\end{align*}
Similarly, the total probability $P$ is bounded as
\[
\Pr\{\overline{A}(\xi_{\ast})\}\leq P\leq\sum_{\xi\in\Gamma}\Pr\{\overline
{A}(\xi)|B(\xi)\}=\sum_{\xi\in\Gamma}p(\xi).
\]
\hfill\hfill$\square$

\subsection{Asymptotic setting}

Given any path $\xi,$ we now assume that decoder gives correct
solutions $\hat{a}(\gamma)=1$ on all previous paths $\gamma<\xi.$
Our next goal is
to estimate the decoding error probability%
\begin{equation}
p(\xi)=\Pr\{y(\xi)<0\} \label{main3}%
\end{equation}
where $y(\xi)$ is a random variable (rv), which satisfies simplified
recalculations (\ref{main2}). Here we begin with the original probability
distribution
\begin{equation}
\Pr\{y_{i}\}=\left\{
\begin{array}
[c]{ll}%
1-p, & \text{if}\;y_{i}=+1,\\
p, & \text{if}\;y_{i}=-1,
\end{array}
\right.  \label{init}%
\end{equation}
where $y_{i}$ are $n$ independent identically distributed \ (i.i.d.) rv that
form the received vector $\mathbf{y.}\smallskip$

\textit{Remark. }Note that the above problem is somewhat similar to that of
``probability density evolution '' researched in iterative algorithms. Namely,
in both algorithms the original rv $y_{i}$\ undergo two different
transformations, similar to (\ref{main2}). However, in our setting these
transformations can also be mixed in an arbitrary (irregular) order that only
depends on a particular path $\xi,$ in general, and on its current symbol
$\xi_{s},$ in particular. $\smallskip$

To simplify this problem, below we estimate $p(\xi)$ using only the first two
moments of variables $y_{i}$ and their descendants. This will be done as follows.\medskip

1. First, note that the blocks $\mathbf{y}^{\prime}$ and $\mathbf{y}%
^{\prime\prime}$ used in (\ref{main1}) always include different
channel bits$.$ Consequently, their descendants
$\mathbf{y}^{\prime}(\,\underline{\xi }\,) $ and
$\mathbf{y}^{\prime\prime}(\,\underline{\xi}\,)$ used in
(\ref{main2}) are also obtained from different channel bits. These
bits are combined in the same operations. Therefore all symbols
$y_{i}(\xi)$ of the vector $\mathbf{y}(\xi)$ are \ i.i.d. rv. This
allows us to use the common notation $y(\xi)$ for any random
variable $y_{i}(\xi)$ obtained on the subpath $\xi.$ \smallskip

2. Let $e(\xi)=\mathsf{E}$ $y(\xi)$ denote the expectation of any
rv $y(\xi).$ Below we study the normalized random variables
\begin{equation}
z(\xi)=y(\xi)/e(\xi), \label{z0}%
\end{equation}
all of which have expectation 1. Our goal is to estimate their variances%
\begin{equation}
\mu(\xi)\overset{\text{def}}{=}\text{$\mathsf{E}$ }(z(\xi)-1)^{2}.
\label{cheb}%
\end{equation}
Then decoding error probability always satisfies Chebyshev's inequality
\begin{equation}
p(\xi)=\Pr\{z(\xi)<0\}\leq\mu(\xi). \label{cheb1}%
\end{equation}

3. To prove Theorem 1, we first consider those left-end paths $\xi$ that pass
through the nodes $\left\{
\genfrac{}{}{0pt}{}{g}{0}%
\right\}  $ with growing $g\geq m^{1/2}.$ \ For any such path, we
show that the corresponding rv $y(\xi)$ satisfies the central
limit theorem as $m\rightarrow\infty.$ This will allow us to
replace Chebyshev's inequality (\ref{cheb1}) by (a stronger)
Gaussian approximation. We will also see that the variance
$\mu(\xi)$ rapidly declines as decoding progresses over the new
paths $\xi$. \ For this reason, we shall still use Chebyshev's
inequality (\ref{cheb1}) on the remaining paths with $g<m^{1/2},$
which will only slightly increase the block error probability $P$
defined in (\ref{tot3}).

\subsection{Recalculation of the variances}

Our next goal is to recalculate the variances $\mu(\xi)$ defined in
(\ref{cheb})$.$\ Let the channel residual $\varepsilon=1-2p$ be fixed.
According to (\ref{init}), original channel outputs $y_{i}$ have the means
$\mathsf{E}y_{i}=\varepsilon,$ in which case rv $z_{i}=$ $y_{i}/\varepsilon$
have the variance
\[
\mu_{0}=\varepsilon^{-2}-1.\smallskip
\]

\begin{lemma}
For any path $\xi=(\,\underline{\xi}\,,\xi_{s}),$ the variance $\mu(\xi)$
satisfies the recursions%
\begin{equation}%
\begin{tabular}
[c]{ll}%
$\mu(\xi)+1=(\mu(\,\underline{\xi}\,)+1)^{2},$ & $\text{if}\;\xi_{s}=0,$%
\end{tabular}
\ \label{t12}%
\end{equation}%
\begin{equation}%
\begin{tabular}
[c]{ll}%
$\qquad\qquad\mu(\xi)=\mu(\,\underline{\xi}\,)/2,$ &
$\text{if}\;\xi
_{s}=1.\medskip$%
\end{tabular}
\ \label{t1}%
\end{equation}
\end{lemma}

\noindent\textit{Proof.} \ First, we need to find the means $e(\xi)$ of rv
$y(\xi)$ to proceed with new variables $z(\xi).$ Here we simply replace all
three rv used in (\ref{main2}) by their expectations. Then for any
$\xi=(\underline{\xi},\xi_{s}),$ the means $e(\xi)$ satisfy the recursion%
\begin{equation}
e(\xi)=\left\{
\begin{array}
[c]{ll}%
e^{2}(\underline{\xi}), & \text{if}\;\xi_{s}=0,\medskip\\
e(\underline{\xi}), & \text{if}\;\xi_{s}=1.\medskip
\end{array}
\right.  \label{e10}%
\end{equation}
Here we also use the fact that vectors
$\mathbf{y}^{\prime}(\underline{\xi})$ and
$\mathbf{y}^{\prime\prime}(\underline{\xi})$ are independent and
have symbols with the same expectation $e(\,\underline{\xi}\,).$
Now we see that the normalized rv $z(\xi)$ satisfy the recursion
\begin{equation}
z(\xi)=\left\{
\begin{array}
[c]{ll}%
z^{\prime}(\underline{\xi})\cdot z^{\prime\prime}(\underline{\xi}), &
\text{if}\;\xi_{s}=0,\medskip\\
z^{\prime}(\underline{\xi})/2+z^{\prime\prime}(\underline{\xi})/2, &
\text{if}\;\xi_{s}=1,\medskip
\end{array}
\medskip\medskip\right.  \label{main5}%
\end{equation}
similarly to (\ref{main2}). By taking $\mathsf{E}z^{2}(\xi)$ we
immediately obtain (\ref{t12}) and
(\ref{t1}).\hfill$\square\medskip{\medskip}$

\textit{Discussion.} Note that\ the means $e(\xi)$ only depend on the Hamming
weight $w(\xi)$ of a binary subpath $\xi.$ Indeed, a subpath $\xi$ has
$s-w(\xi)$ zero symbols $\xi_{i}.$ According to (\ref{e10}), the original
expectation $\mathsf{E}(y)$ of rv $y_{i}$ is squared $s-w(\xi)$ times and is
left unchanged $w(\xi)$ times. Therefore
\begin{equation}
e(\xi)=\varepsilon^{2^{s-w(\xi)}}. \label{e11}%
\end{equation}
By contrast, equalities (\ref{t12}) and (\ref{t1}) show that variance $\mu
(\xi)$ depends on positions of all ones in vector $\xi$. Thus, direct
(nonrecurrent) calculations of $\mu(\xi)$\ become more involved. In Lemma
\ref{lm:mu}, we will see that even the simplest paths give rather bulky
expressions (\ref{mu1}) for $\mu(\xi).$ For this reason, we use a different
approach. Namely, in the next section we find the paths $\xi$ that maximize
$\mu(\xi)$.

\subsection{The weakest paths}

\noindent\textit{Preliminary discussion. \ }Consider a channel
with crossover error probability\textit{\ }$(1-\varepsilon)/2$ and
residual $\varepsilon.$ Initially, rv $z(\xi)$ have the variance
$\mu_{0}=\varepsilon^{-2}-1$ and always satisfy inequality
$\mu(\xi)>0,$ by definition (\ref{cheb}). According to
(\ref{t12}), $\mu(\,\underline{\xi}\,)+1$ is always squared when a
path $\underline{\xi}$ is appended by $\xi_{s}=0.$ Thus, moving
from a code $\left\{
\genfrac{}{}{0pt}{}{m}{r}%
\right\}  $ to its left descendant $\left\{
\genfrac{}{}{0pt}{}{m-1}{r-1}%
\right\}  $ is equivalent to the replacement of the original residual
$\varepsilon$ by its square $\varepsilon^{2}.$ In other words, any left-hand
movement makes the descendant channel noisier. For small $\mu(\,\underline
{\xi}\,)\approx0$ (very high quality channel), squaring $\mu(\,\underline{\xi
}\,)+1$ is almost insignificant. However, it becomes more substantial as
$\mu(\,\underline{\xi}\,)$ grows.

By contrast, $\mu(\,\underline{\xi}\,)$ is always cut in half$,$
when $\xi _{s}=1.$ In general, any right-hand movement makes the
descendant channel less noisy. For example, we obtain
$\mu_{0}\approx\varepsilon^{-2}$ on (bad) channels with small
residual $\varepsilon.$ Then performing the right step, the
recursion replaces this residual $\varepsilon$ with the quantity
almost equal to $\varepsilon\sqrt{2}.$ Therefore our first
conclusion is that variance $\mu(\xi)$ increases if $\xi_{s}=1$ is
replaced by $\xi_{s}=0$.\bigskip

\textit{Neighboring paths. }Our next step is to consider two ``equally
balanced'' \ movements. Namely, \ in (\ref{loop}) below, we consider two
subpaths $\xi_{-}$ and $\xi_{+}$ \ of length $s$ that have the same prefix
$\underline{\xi}\,$of length $s-2$ but diverge in the last two positions as follows%

\begin{equation}
\left\{
\begin{array}
[c]{l}%
\xi_{-}=(\underline{\xi},0,1),\ \ \ \ \ \ \ \ \ \,\ \,\nearrow\nwarrow\ \\
\ \ \ \ \ \ \ \ \ \ \ \ \ \ \ \ \ \ \ \ \ \ \ \ \ \ \ \ \ \,\ \,\ ^{0}%
\nwarrow\nearrow\,^{1}\ \\
\xi_{+}=(\underline{\xi},1,0).\ \ \ \ \ \underline{\xi}=\xi_{1},...,\xi_{s-2}%
\end{array}
\right.  \label{loop}%
\end{equation}
We say that $\xi_{-}$ and $\xi_{+}$ are left and right \textsf{neighbors}, correspondingly.\smallskip

\begin{lemma}
\label{lm:nei1} Any two neighbors $\xi_{-}$ and $\xi_{+}$ satisfy inequality%
\begin{equation}
\mu(\xi_{-})\geq\mu(\xi_{+}). \label{in1}%
\end{equation}
\smallskip
\end{lemma}

\noindent\textit{Proof.} Let $\mu\mathbf{(}\underline{\xi})=\tau.$ Then we use
recursive equations (\ref{t12}) and (\ref{t1}), which give
\[
\mu(\xi_{-})=\tau^{2}/2+\tau,\quad\mu(\xi_{+})=\tau^{2}/4+\tau.
\]
Therefore (\ref{in1}) holds$.$\hfill$\square\medskip{\medskip}$

\textit{The weakest paths. }Now we see that any path $\xi$ that
includes two adjacent symbols $(1,0)$ increases its $\mu(\xi)$
after permutation ($1,0)\Rightarrow(0,1).$ In this case, we say
that this path $\xi$ becomes \textsf{weaker}. From now on, let
$\Gamma$ \ be the complete set of \ $k$ extended paths $\xi.$
Also, let $\Gamma^{g}$ be the subset\ of all left-end paths $\xi$
that enter the node $\left\{
\genfrac{}{}{0pt}{}{g}{0}%
\right\}  $ and $\Gamma_{0}$ be the subset of the right-end paths. Given any
subset $I\subseteq\Gamma,$ we now say that $\xi_{\ast}(I)$ is the weakest path
in $I$ if
\[
\mu\mathbf{(}\xi_{\ast}(I))=\max\{\mu\mathbf{(}\xi)|\xi\in I\}.
\]
Then we have the following.${\medskip}$

\begin{lemma}
\label{lm:path}The weakest path on the full set $\Gamma$ of all $k$ paths is
the leftmost path (\ref{w1}). More generally, for any $g\in\lbrack1,m-r],$ the
weakest path on the subset $\Gamma^{g}$ is its leftmost path%
\begin{equation}
\xi_{\ast}^{g}=(0^{r-1},1^{m-r-g},0,1^{g}). \label{w2}%
\end{equation}
$\smallskip$
\end{lemma}

\noindent\textit{Proof.} \ First, note that on all left-end paths
$\xi$, the variances $\mu\mathbf{(}\xi)$ are calculated after $m$
steps, at the same node $\left\{
\genfrac{}{}{0pt}{}{0}{0}%
\right\}  .$ By contrast, all right-end paths $\xi$ end at
different nodes $\left\{
\genfrac{}{}{0pt}{}{h}{h}%
\right\};$   therefore their variances $\mu(\xi)$ are found after
$m-h$ steps. To use Lemma \ref{lm:nei1}, we consider an extended
right-end path $\xi^{e}=(\xi,0^{h})$ obtained by adding $h$ zeros.
Then we have inequality $\mu(\xi^{e})>\mu(\xi),$ since the
variance $\mu$ increases after zeros are added. Despite this fact,
below we prove that $\xi_{\ast}$ from (\ref{w1}) and
$\xi_{\ast}^{g}$ from (\ref{w2}) still represent the weakest
paths, even after this extension.

Indeed, now all paths have the same length $m$ and the same weight
$m-r,$ so we can apply Lemma \ref{lm:nei1}. Recall that each path
$\xi\in\Gamma^{g}$ ends with the same suffix $0,1^{g}$. In this
case, $\xi_{\ast}^{g}$ is the leftmost path on $\Gamma^{g}$. By
Lemma \ref{lm:nei1}, $\xi_{\ast}^{g}$ maximizes the variance
$\mu(\xi)$ over all $\xi\in\Gamma^{g}.$ Finally, note that
$\xi_{\ast}$ \ is the leftmost path on the total set $\Gamma$
since all $r$ zeros form its prefix $0^{r}$. Thus, $\xi_{\ast}$ is
the weakest \ path. \hfill$\square$

\section{Threshold of algorithm $\Psi_{\,r}^{m}$}

Now we find the variances $\mu\mathbf{(}\xi_{\ast})$ and $\mu\mathbf{(}%
\xi_{\ast}^{g})$ for the weakest paths $\xi_{\ast}$ and $\xi_{\ast}%
^{g}.{\medskip}$

\begin{lemma}
\label{lm:mu} For crossover error probability $(1-\varepsilon)/2$, the weakest
paths $\xi_{\ast}$ and $\xi_{\ast}^{g}$ give the variances%
\begin{equation}
\mu\mathbf{(}\xi_{\ast})=2^{-(m-r)}(\varepsilon^{-2^{r+1}}-1), \label{t0}%
\end{equation}%
\begin{equation}
\mu\mathbf{(}\xi_{\ast}^{g})=2^{-g}(((\varepsilon^{-2^{r}}-1)2^{r+g-m}%
+1)^{2}-1). \label{mu1}%
\end{equation}
$\smallskip$
\end{lemma}

\noindent\textit{Proof. }\ Consider the weakest path $\xi_{\ast}$
from (\ref{w1}). The recursion (\ref{t12}) begins with the
original quantity $\mu(\xi)+1=\varepsilon^{-2}.$ After completing
$r$ left steps $\underline {\xi}=\ 0^{r},$ the result is
\begin{equation}
\mu_{r}(\underline{\xi}\,)+1=\varepsilon^{-2^{r+1}}. \label{t2}%
\end{equation}
Then we proceed with $m-r$ right steps, each of which cuts $\mu_{r}%
\mathbf{(}\,\underline{\xi}\,)$ in half according to (\ref{t1}).
Thus, we obtain equality (\ref{t0}). Formula (\ref{mu1}) follows
from representation (\ref{w2}) \ in a similar (though slightly
longer) way. {\medskip} \hfill$\square$

Lemma \ref{lm:mu} allows us to use Chebyshev's inequality
\begin{equation}
p(\xi)\leq\mu\mathbf{(}\xi)\leq\mu\mathbf{(}\xi_{\ast}) \label{cheb2}%
\end{equation}
for any path $\xi.$ \ However, this bound is rather loose and
insufficient to prove Theorem 1. Therefore we improve this
estimate, separating all paths into two different sets. Namely,
let $\Gamma^{\ast}$ \ be the subset\ of all left-end paths that
enter the node $\left\{
\genfrac{}{}{0pt}{}{g}{0}%
\right\}  $ with $g\geq m^{1/2}.$

We will use the fact that any path $\xi\in\Gamma^{\ast}$ satisfies the central
limit theorem as $m$ grows. However, we still use Chebyshev's inequality on
the complementary subset $\Gamma\setminus\Gamma^{\ast}.$ In doing so, we take
$\varepsilon$ equal to the $\varepsilon_{r}$ from Theorem 1:
\begin{equation}
\varepsilon_{r}=(d^{-1}2r\ln m)^{1/2^{r+1}},\quad m\rightarrow\infty.
\label{set}%
\end{equation}

\begin{theorem}
\label{lm:fi}For RM codes with $m\rightarrow\infty$ and fixed order $r$ used
on a binary channel with crossover error probability $(1-\varepsilon_{r})/2,$
algorithm $\Psi_{\,r}^{m}$ gives on a path $\xi$ the asymptotic bit error
rate
\begin{equation}
p(\xi)\lesssim m^{-r}/\sqrt{4\pi r\ln m},\quad m\rightarrow\infty
\label{gauss1}%
\end{equation}
with asymptotic equality on the weakest path
$\xi_{\ast}$.\noindent \medskip
\end{theorem}

\textit{Proof. }\ According to (\ref{md2}), any left-end path $\xi$ gives the
rv $y(\xi),$ which is the sum of $\ 2^{g(\xi)}$ i.i.d. limited rv $y_{i}%
(\xi).$ For $\xi\in\Gamma^{\ast},$ this number grows as $2^{\sqrt{m}}$ or
faster as $m\rightarrow\infty.$ In this case, the normalized rv $z(\xi)$
satisfies the central limit theorem and its probability density function (pdf)
tends to the Gaussian distribution $\mathcal{N}(1,\mu\mathbf{(}\xi)).$

According to Lemmas \ref{lm:path} and \ref{lm:mu}, the weakest path $\xi
_{\ast}$ gives the maximum variance $\mu\mathbf{(}\xi_{\ast}).$ In particular,
for $\varepsilon=\varepsilon_{r}$ equality (\ref{t0}) gives
\begin{equation}
\mu\mathbf{(}\xi_{\ast})=(2r\ln m)^{-1}-2^{r-m}. \label{small}%
\end{equation}
Using Gaussian distribution $\mathcal{N}(1,\mu\mathbf{(}\xi_{\ast}))$ to
approximate $p(\xi_{\ast}),$ we take $\mu^{-1/2}\mathbf{(}\xi_{\ast})$
standard deviations and obtain (see also Remark 1 following the proof)
\begin{equation}
p(\xi_{\ast})\sim Q(\mu^{-1/2}\mathbf{(}\xi_{\ast})),\quad m\rightarrow\infty.
\label{gauss}%
\end{equation}
Here we also use the asymptotic%
\[%
\begin{tabular}
[c]{ll}%
$Q(x)$ & $\overset{\text{def}}{=}\int_{x}^{\infty}\exp\{-x^{2}/2\}dx/\sqrt
{2\pi}\medskip$\\
& $\sim\exp\{-x^{2}/2\}/(x\sqrt{2\pi})$%
\end{tabular}
\ \ \
\]
valid for large $x.$ This yields asymptotic equality for $p(\xi_{\ast})$ in
(\ref{gauss1}). For any other path $\xi\in\Gamma^{\ast},$ $z(\xi)$ \ is
approximated by the normal rv with a smaller variance $\mu\mathbf{(}\xi
)<\mu\mathbf{(}\xi_{\ast}).$ Therefore we use inequality in (\ref{gauss1}):
\begin{equation}
p(\xi)<Q(\mu^{-1/2}\mathbf{(}\xi_{\ast})). \label{gauss2}%
\end{equation}
Finally, consider any path $\xi$ with $g<m^{1/2}.$ In this case,
we directly estimate the asymptotics of
$\mu\mathbf{(}\xi_{\ast}^{g}).$ Namely, we use the substitution
$\varepsilon=\varepsilon_{r}$ in (\ref{mu1}), which gives a useful
estimate:%
\begin{equation}
\mu\mathbf{(}\xi_{\ast}^{g})\sim\left\{
\begin{array}
[c]{ll}%
2^{-m-r+g}(2r\ln m)^{-1}, & \text{if}\;g>\frac{m-r}{2}+\ln m,\smallskip\\
2^{-(m-r-2)/2}(2r\ln m)^{-1/2}, & \text{if}\;g<\frac{m-r}{2}-\ln m,
\end{array}
\right.  \label{mu-low}%
\end{equation}

$\smallskip$ Thus, we see that for all $g<m^{1/2},$ variances $\mu
\mathbf{(}\xi_{\ast}^{g})$ have the same asymptotics and decline
exponentially in $m,$ as opposed to the weakest estimate
(\ref{small}). Then we have
\[
p(\xi)\leq\mu\mathbf{(}\xi)\leq\mu\mathbf{(}\xi_{\ast}^{g})<2^{-(m-r)/2},
\]
which also satisfies (\ref{gauss1}) as $m\rightarrow\infty.$\hfill
$\square\medskip$

\textit{Discussion.}

1. Considering approximation (\ref{gauss}) for a general path $\xi,$ we arrive
at the estimate
\begin{equation}
p(\xi)\sim Q(\mu^{-1/2}\mathbf{(}\xi)),\quad m\rightarrow\infty.
\label{gauss5}%
\end{equation}
According to Theorem XVI.7.1 from \cite{fel}, this approximation is valid if
the number of standard deviations $\mu^{-1/2}\mathbf{(}\xi)$ is small relative
to the number $2^{g\mathbf{(}\xi)}$ of rv $z_{i}(\xi)$ in the sum $z(\xi):$
\begin{equation}
\mu^{-1/2}\mathbf{(}\xi)=o(2^{g\mathbf{(}\xi)/6}). \label{gauss3}%
\end{equation}
In particular, we can use (\ref{gauss}) for the path $\xi_{\ast},$
since (\ref{small}) gives
\[
\mu^{-1/2}\mathbf{(}\xi_{\ast})\sim(2r\ln m)^{1/2}=o(2^{\sqrt{m}/6}).
\]
$\smallskip$ 2. Note that for $\varepsilon=\varepsilon_{r},$ variance
$\mu\mathbf{(}\xi_{\ast}^{g})$ in (\ref{mu-low}) declines exponentially as $g$
moves away from $m-r$. On the other hand, \ we can satisfy\ asymptotic
condition (\ref{gauss3}) for any path $\xi\in\Gamma^{\ast},$ if $\mu
^{-1/2}\mathbf{(}\xi)$ in (\ref{gauss3}) is replaced with
parameter\footnote{Here we take any positive polynomial $\mathsf{poly(m)}$ of
a fixed degree as $m\rightarrow\infty.$}
\[
\tau^{-1/2}\mathbf{(}\xi)=\min\{\mu^{-1/2}\mathbf{(}\xi),\mathsf{poly(}m)\}
\]
as $m\rightarrow\infty.$ We then use inequality
\[
p\mathbf{(}\xi)\leq Q(\tau^{-1/2}\mathbf{(}\xi))
\]
valid for any $\xi\in\Gamma^{\ast},$ instead of the weaker inequality
(\ref{gauss2}). Thus, the bounds on probabilities $p(\xi)$ rapidly decline as
$g$ moves away from $m-r,$ and the total block error rate $P$ also satisfies
the same asymptotic bound (\ref{gauss1})$.\smallskip$

3. Note that\ the same minimum residual $\varepsilon_{r}$ can also be used for
majority decoding. Indeed, both the majority and the recursive algorithms are
identical on the weakest path $\xi_{\ast}$. Namely, both algorithms first
estimate the product of $2^{r}$ channel symbols and then combine $2^{m-r}$
different estimates in (\ref{md2}). However, a substantial difference between
the two algorithms is that recursive decoding uses the previous estimates to
process any other path $\xi$. Because of this, the algorithm outperforms
majority decoding in both the complexity and BER $p(\xi)$ for any $\xi\neq
\xi_{\ast}.\smallskip$

4. Theorem \ref{lm:fi} almost entirely carries over to any $\varepsilon
>\varepsilon_{r}$. Namely, we use the normal pdf $\mathcal{N}(1,\mu
\mathbf{(}\xi))$ for any $\xi\in\Gamma^{\ast}$ . Here any variance
$\mu\mathbf{(}\xi)$ declines as $\varepsilon$ grows. Therefore we
can always employ inequality (\ref{gauss2}), by taking the
\textit{maximum possible} variance $\mu\mathbf{(}\xi_{\ast})$
obtained in (\ref{small}). On the other hand, asymptotic equality
(\ref{gauss}) becomes invalid as $\varepsilon$ grows.

In this case, tighter bounds (say, the Chernoff bound) must be used on
$\xi_{\ast}.$ However, in this case, we also need to extend the second-order
analysis of Lemma \ref{lm:nei1} to exponential moments. Such an approach can
also give the asymptotic error probability $p(\xi)$ for any $\varepsilon>$
$\varepsilon_{r}.$ However, finding the bounds on $p(\xi)$ is an important
issue still open to date. $\smallskip$

5. It can be readily proven that for sufficiently large $\varepsilon
>2^{-(m-r-g)/2^{r}},$ the variance $\mu\mathbf{(}\xi_{\ast}^{g})$ becomes
independent of \ $g,$ similar to the estimates obtained in the second line of
(\ref{mu-low}). More generally, more and more paths yield almost equal
contributions to the block error rate as $\varepsilon$ grows. This is due to
the fact that the neighboring paths exhibit similar performance on
sufficiently good channels. \medskip

Now Theorem 1 directly follows from Theorem \ref{lm:fi}.

\noindent\textit{Proof of Theorem 1. }\ Consider a channel with crossover
probability $p=(1-\varepsilon_{r})/2$ \ for $m\rightarrow\infty.$ The output
block error probability $P$ of the algorithm $\Psi_{\,r}^{m}$ has the order
$\ $at most $kp(\xi_{\ast}),$ where $k$ is the number of information symbols.
This number has polynomial order of $\left(
\genfrac{}{}{0pt}{}{m}{r}%
\right)  .$ On the other hand, \ formula (\ref{gauss1}) shows that
$p(\xi_{\ast})$ declines faster than $k^{-1}$ for any $\varepsilon
\geq\varepsilon_{r}.$ As a result, $P\rightarrow0.$

Next, we note that the error patterns of weight $pn$ or less occur with a
probability
\[
\sum_{i=0}^{pn}\left(  _{i}^{n}\right)  p^{i}(1-p)^{n-i}\rightarrow Q(0)=1/2.
\]
Since $P\rightarrow0,$ the above argument shows that decoder fails to decode
only a vanishing fraction of error patterns of weight $pn$ or less$.$

Next, we need to prove that $\Psi_{\,r}^{m}$ fails to correct nonvanishing
fraction of errors of weight \ $n/2$ or less. \ In proving this, consider a
higher crossover probability $p_{1}=(1-\varepsilon)/2,$ where%
\[
\varepsilon=\varepsilon_{r}(\ln m)^{-1/2^{r}}.
\]
For this $\varepsilon,$ our estimates (\ref{t0}) and (\ref{gauss}) show that
$\mu\mathbf{(}\xi_{\ast})\rightarrow0$ and $p(\xi_{\ast})\rightarrow1/2.$
Also, according to (\ref{tot3}), $P>p(\xi_{\ast}).$ On the other hand, the
central limit theorem shows that the errors of weight $n/2$ or more still
occur with a vanishing probability
\[
\sum_{i=n/2}^{n}\left(  _{i}^{n}\right)  p_{1}^{i}(1-p_{1})^{n-i}%
\rightarrow0.
\]
Thus, we see that $\Psi_{\,r}^{m}$ necessarily fails on the
weights \ $n/2$ or less, since the weights over $n/2$ still give a
vanishing contribution to the nonvanishing error rate
$p(\xi_{\ast})$. \hfill$\square$

\section{Threshold of algorithm $\Phi_{\,r}^{m}$}

\noindent Before proceeding with a proof of Theorem \ref{th:1-1}, we summarize
three important points that will be used below to evaluate the threshold of
the algorithm $\Phi_{\,r}^{m}$. {\smallskip}

\textbf{1.} The received rv $y_{i}$, all intermediate recalculations
(\ref{main}), and end decodings on the right-end paths $\xi$ are identical in
both algorithms $\Phi_{\,r}^{m}$ and $\Psi_{\,r}^{m}.$

By contrast, any left-end path $\xi=\underline{\xi},1^{g}$ first arrives at
some biorthogonal code $\left\{
\genfrac{}{}{0pt}{}{g+1}{1}%
\right\}  $ of length $l=2^{g+1}$ and is then followed by the suffix $1^{g}.$
Also, let $c_{s}^{g}$ denote the $s$th codeword of $\left\{
\genfrac{}{}{0pt}{}{g+1}{1}%
\right\}  ,$ where $s=1,...,2l.$ Here we also assume that the first two
codewords form the repetition code:
\[
c_{1}^{g}=1^{l},\quad c_{2}^{g}=-c_{1}^{g}.
\]
For each $c_{s}^{g},$ define its support $I_{s}^{g}$ \ as the subset of
positions that have symbols $\ -1$. Here $2l-2$ codewords with $s>2$ have
support of the same size $|I_{s}^{g}|=2^{g},$ whereas $|I_{2}^{g}|=2^{g+1}.$
Also, below $a(\xi)$ denotes any information symbol associated with a path
$\xi$. {\smallskip}

\textbf{2.} Let the all-one codeword $1^{n}$ be transmitted and $\mathbf{y}$
be received. Consider the vector $\mathbf{y}(\underline{\xi})$ obtained on
some left-end path $\underline{\xi}$ that ends at the node $\left\{
\genfrac{}{}{0pt}{}{g+1}{1}%
\right\}  .$ By definition of MD decoding, $\mathbf{y}(\underline{\xi})$ is
incorrectly decoded into any $c\neq c_{1}^{g}$ with probability
\[
P(\xi)=\Pr\left\{  \cup_{s=2}^{2l}\,\,\Omega_{s}(\xi)\right\}
\]
where
\begin{equation}
\Omega_{s}(\xi)=\left\{  \mathbf{y}:\sum_{i\in I_{s}^{g}}y_{i}(\underline{\xi
})<0\right\}  .\label{main4}%
\end{equation}
In our probabilistic setting, each event $\Omega_{s}(\xi)$ is completely
defined by the symbols $y_{i}(\underline{\xi}),$ which are i.i.d.
rv$.${\smallskip}

\textbf{3.} Recall that Lemma \ref{lm:prob0} is ``algorithm-independent'' and
therefore is \ left intact in $\Phi_{\,r}^{m}$. Namely, we again consider the
events $A(\xi)$ and $B(\xi)$ from (\ref{events}). Similarly to (\ref{tot4}),
we assume that all preceding decodings are correct and replace the
unconditional error probability $P(\xi)$ with its conditional counterpart
\[
p(\xi)=\Pr\{\overline{A}(\xi)\,|\,B(\xi)\}.
\]
This probability satisfies the \ bounds%
\begin{equation}
\Pr\Omega_{s}(\xi)\leq p(\xi)\leq\sum_{s=2}^{2l}\Pr\Omega_{s}(\xi).
\label{b-fi}%
\end{equation}
Here we take the probability of incorrect decoding into any single codeword
$c_{s}^{g}$ as a lower bound (in fact, below we choose $s>2$), and the union
bound as its upper counterpart.\medskip

Now we take parameters
\[
\tilde{\varepsilon}_{r}=(cm2^{r-m})^{1/2^{r}},\quad c>\ln4,\quad c^{\prime
}=c/2-\ln2.
\]

\begin{theorem}
For RM codes with $m\rightarrow\infty$ and fixed order $r$ used on a binary
channel with crossover error probability $(1-\tilde{\varepsilon}_{r})/2,$
algorithm $\Phi_{\,r}^{m}$ gives for any path $\xi$ a vanishing bit error rate%
\begin{equation}
p(\xi)<\max\{e^{-c^{\prime}m},2^{-(m-r)/2+m^{1/2}}\}. \label{log2}%
\end{equation}
${\smallskip}$
\end{theorem}

\noindent\textit{Proof. }\ Consider any left-end path $\underline{\xi}$ that
ends at the node $\left\{
\genfrac{}{}{0pt}{}{g+1}{1}%
\right\}  .$ For any vector $\mathbf{y}(\underline{\xi})$ and any subset $I$
\ of $2^{g}$ positions, define the sum
\[
y_{I}(\xi)\overset{\text{def}}{=}\sum_{i\in I}y_{i}(\underline{\xi}).
\]
Here $y_{i}(\underline{\xi})$ form $2^{g}$ i.i.d. rv. Thus, the sum $y_{I}%
(\xi)$ has the same pdf for any $I$. In turn, this allows us to remove index
$I$ from $y_{I}(\xi)$ and use common notation $y(\xi)=y_{I}(\xi).$ Then we
rewrite bounds (\ref{b-fi}) as%
\[
\Pr\{y(\xi)\leq0\}<p(\xi)\leq(2l-1)\Pr\{y(\xi)\leq0\}.
\]
Equivalently, we use the normalized rv $z(\xi)=y(\xi)/\mathsf{E}y(\xi)$ with
expectation 1 and rewrite the latter bounds as%
\begin{equation}
\Pr\{z(\xi)\leq0\}<p(\xi)\leq(2l-1)\Pr\{z(\xi)\leq0\}.\label{main6}%
\end{equation}
Similarly to the proof of Theorem \ref{lm:fi}, note that the sum $z(\xi)$ also
satisfies the central limit theorem for any $g\geq m^{1/2}$ and has pdf that
tends to $\mathcal{N}(1,\mu(\xi))$ as $m\rightarrow\infty.$ Thus, we see that
$p(\xi)$ depends only on the variance $\mu(\xi)\mathbf{\ }$ obtained on the
sum $z(\xi)$ of i.i.d. rv. This variance can be found using calculations
identical to those performed in Lemmas \ref{lm:nei1} to \ref{lm:mu}. In
particular, for any $g$ we can invoke the proof of Lemma \ref{lm:path}, which
shows that $\mu(\xi)$ achieves its maximum on the leftmost path%
\[
\xi_{\ast}=(0^{r-1},1^{m-r}).
\]
Similarly to (\ref{t0}), we then find%
\begin{equation}
\mu\mathbf{(}\xi_{\ast})=2^{-(m-r)}(\varepsilon^{-2^{r}}-1).\label{t14}%
\end{equation}
Direct substitution $\varepsilon=\tilde{\varepsilon}_{r}$ in (\ref{t14})
gives
\[
\mu\mathbf{(}\xi_{\ast})=(cm)^{-1}-2^{r-m}.
\]
Now we almost repeat the proof of Theorem \ref{lm:fi}. For the first path
$\xi_{\ast},$ we employ Gaussian approximation
\[
\Pr\{z(\xi_{\ast})\leq0\}\sim Q(\mu^{-1/2}\mathbf{(}\xi_{\ast}))\sim
e^{-cm/2}(2\pi cm)^{-1/2}%
\]
as $m\rightarrow\infty.$ For maximum $l=2^{m-r+1},$ the latter inequality and
(\ref{main6}) give the upper bound
\begin{align}
p(\xi_{\ast}) &  \leq2^{m-r+2}e^{-cm/2}(2\pi cm)^{-1/2}\label{t15}\\
&  <e^{-c^{\prime}m}.\nonumber
\end{align}
Also,
\begin{equation}
p(\xi_{\ast})\geq Q(\mu^{-1/2}\mathbf{(}\xi_{\ast})).\label{lo1}%
\end{equation}
For any other path $\xi$ with $g\geq m^{1/2},$ we can use the same estimates
in (\ref{main6}) due to the inequalities $\mu\mathbf{(}\xi)<\mu\mathbf{(}%
\xi_{\ast})$ and $l\leq2^{m-r}.$

Finally, consider any path $\xi$ with $g\leq m^{1/2}.$ In this case, we use
Chebyshev's inequality instead of Gaussian approximation. Again, for any node
$\left\{
\genfrac{}{}{0pt}{}{g+1}{1}%
\right\}  ,$ we can consider its leftmost path
\[
\xi_{1}^{g}=(0^{r-2},1^{m-r-g},0,1^{g}).
\]
Similarly to our previous calculations in (\ref{t0}) and (\ref{mu1}), it can
be verified that%
\[
\mu\mathbf{(}\xi_{1}^{g})=2^{-g}(((\varepsilon^{-2^{r-1}}-1)2^{r+g-m}%
+1)^{2}-1).
\]
Then for small $g\leq m^{1/2}$, substitution $\varepsilon=\tilde{\varepsilon
}_{r}$ gives the equality
\begin{equation}
\mu\mathbf{(}\xi_{1}^{g})\sim2^{-(m-r-2)/2}(cm)^{-1/2},\quad
m\rightarrow \infty.\label{cheb7}
\end{equation}
Thus, we obtain Chebyshev's inequality in the form
\begin{equation}
p(\xi_{1}^{g})\leq\mu\mathbf{(}\xi_{1}^{g})\cdot
2^{g+2}<2^{-(m-r)/2+m^{1/2}}
\label{cheb3}%
\end{equation}
and complete the proof, since bound (\ref{log2}) combines both
estimates (\ref{t15}) and
(\ref{cheb3}).\hfill\hfill$\square{\smallskip}$

\textit{Proof of Theorem 2. }\ We repeat the proof of Theorem 1 almost
entirely. Consider a channel with crossover probability $(1-\tilde
{\varepsilon}_{r})/2$ \ as $m\rightarrow\infty.$ The output block error
probability $P$ of the algorithm $\Phi_{\,r}^{m}$ satisfies the estimate
$P\leq k\max p(\xi),$ where the number $k$ of different paths $\xi$ is bounded
by $\left(
\genfrac{}{}{0pt}{}{m}{r}%
\right)  .$ Formula (\ref{log2}) shows that all $p(\xi)$ decline exponentially
in $m.$ As a result, we obtain asymptotic estimate $P\rightarrow0.$ On the
other hand, the error patterns of weight $pn$ or less occur with a total
probability that tends to $1/2.$ So decoder fails to decode only a vanishing
fraction of these error patterns.

Now we take a smaller residual%
\[
\varepsilon=\tilde{\varepsilon}_{r}/m^{1/2^{r-1}}%
\]
and consider a channel with crossover probability $(1-\varepsilon)/2.$ Then
direct substitution of $\varepsilon$ in (\ref{t14}) gives $\mu\mathbf{(}%
\xi_{\ast})\rightarrow\infty$. Then formula (\ref{lo1}) shows that the
decoding block error rate is
\[
P\geq Q(\mu^{-1/2}(\xi_{\ast}))\rightarrow1/2.
\]
Note also that errors of weight $n/2$ or more occur with vanishing
probability. Thus, $\Psi_{\,r}^{m}$ fails on errors of weight \
$n/2$ or less. \hfill$\square{\smallskip\smallskip}$

\noindent\textit{Discussion. }

The proofs of Theorems 1 and 2 also reveal the main shortcoming of our
probabilistic technique, which employs rather loose estimates for
probabilities $p(\xi).$ Indeed, the first two moments of the random variables
$y(\xi)$ give tight approximation only for Gaussian rv. By contrast, error
probabilities $p(\xi)$ slowly decline as $2^{-(m-r)/2},$ whenever Chebyshev's
inequality (\ref{cheb3}) is applied for small parameters $g<m^{1/2}.$ As a
result, we can obtain a vanishing block error rate only if
\[
k=o(2^{-(m-r)/2}).
\]
This is the case of RM codes of fixed order $r.$\smallskip

By contrast, the number of information symbols $k$ is linear in $n$ for RM
codes of fixed rate $R\in(0,1).$ This fact does not allow us to extend
Theorems 1 and 2 for nonvanishing code rates. More sophisticated arguments -
that include the moments $\mathsf{E}$$z^{s}(\xi)$ of \ an arbitrary order $s$
- can be developed in this case. The end result of this study is that
recursive decoding of RM codes $\left\{
\genfrac{}{}{0pt}{}{m}{r}%
\right\}  $ of fixed rate $R$ achieves the error-correcting threshold
\[
\delta=(d\ln d)/2.
\]
This increases $\ln d$ times the threshold of bounded distance decoding.
However, the overall analysis becomes more involved and is beyond the scope of
this paper.

\section{Further enhancements and open problems}

Now consider an infinite sequence of optimal binary codes of a low code rate
$R$ used on a channel with high crossover error probability $p=(1-\varepsilon
)/2$. According to the Shannon coding theorem, ML decoding of such a sequence
gives a vanishing block error probability if \
\[
p<H^{-1}(1-R),
\]
where $H^{-1}$ is the inverse (binary) entropy function. Note that for
$R\rightarrow0:$%
\[
1-2H^{-1}(1-R)\sim\sqrt{R\ln4}.
\]
Correspondingly, a vanishing block error probability is obtained for any
residual
\begin{equation}
\varepsilon_{\text{opt}}\sim\sqrt{cR},\quad c>\ln4.\label{opt}%
\end{equation}
Next, recall that RM codes $\{_{\,r}^{m}\}$ of fixed order $r$ have code rate
\[
R\sim m^{r}n^{-1}(r!)^{-1}.
\]
For this rate, ML decoding of optimal codes gives
\begin{equation}
\varepsilon_{\text{opt}}\sim
(cm^{r})^{1/2}n^{-1/2}(r!)^{-1/2},\quad m\rightarrow
\infty.\label{ord-ml}%
\end{equation}
Thus, we see that optimal codes give  approximately the same residual order
(\ref{ord-ml}) as the former order (\ref{ml}) derived in \cite{sid} for RM
codes $\{_{\,r}^{m}\}$. \ In other words, RM codes of low rate $R$ can achieve
nearly optimum performance for ML decoding. By contrast, low-complexity
algorithm $\Phi_{\,r}^{m}$ has a substantially higher residual that has the
order of $(m/n)^{1/2^{r}}$. This performance gap shows that further advances
are needed for the algorithm $\Phi_{\,r}^{m}.$ The main problem here is
whether possible improvements can be coupled with low complexity order of
$n\log n.$

The performance gap becomes even more noticeable, if \ a binary symmetric
channel is considered as a ``hard-decision'' image of an AWGN channel. Indeed,
let the input symbols $\pm1$ be transmitted over a channel with the additive
white Gaussian noise $\mathcal{N}(0,\sigma^{2}).$ For code sequences of rate
$R\rightarrow0,$ we wish to obtain a vanishing block error probability when
$\sigma\rightarrow\infty.$ In this case, the transmitted symbols $\pm1$ are
interchanged with very high crossover probability $Q(1/\sigma),$ which gives
residual
\begin{equation}
\varepsilon\sim1-2Q(1/\sigma)\sim\sigma^{-1}\sqrt{2/\pi}.\label{errh}%
\end{equation}
Thus,
\[
\varepsilon^{-2}\sim\pi\sigma^{2}/2
\]
serves (up to a small factor of $\pi/2)$ as a measure of noise power
$\sigma^{2}.$ \ In particular, ML decoding operates at the above residual
$\varepsilon_{\text{opt}}$ from (\ref{ord-ml}) and can withstand noise power
$\sigma^{2}$ of order up to $nm^{-r}.$

By contrast, algorithm $\Phi_{\,r}^{m}$ can successfully operate only when
noise power $\sigma^{2}$ has the lower order of $(n/m)^{1/2^{r-1}}.$
\ Similarly, algorithm $\Psi_{\,r}^{m}$ is efficient when $\sigma^{2}$ is
further reduced to the order of $(n/m)^{1/2^{r}}$. \ Therefore for long RM
codes, algorithm $\Phi_{\,r}^{m}$ can increase $(n/m)^{1/2^{r}}$ times the
noise power that can be sustained using the algorithm $\Psi_{\,r}^{m}$ or
majority decoding. However, performance of $\Phi_{\,r}^{m}$ also degrades for
longer blocks when compared to optimum decoding, though this effect is slower
in $\Phi_{\,r}^{m}$ than in other low-complexity algorithms known for RM codes.

For moderate lengths, this relative degradation is less pronounced, and
algorithm $\Phi_{\,r}^{m}$ achieves better performance. In particular, some
simulation results are presented in Fig.~\ref{fig2} to Fig.~\ref{fig4} for RM
codes $\{_{2}^{7}\}$, $\{_{2}^{8}\}$, and $\{_{3}^{8}\}$, respectively. On the
horizontal axis, we plot both input parameters - the signal-to noise ratio
$(2R\sigma^{2})^{-1}$ of an AWGN channel and the crossover error probability
$Q(1/\sigma)$ of the corresponding binary channel. The output code word error
rates (WER) of algorithms $\Psi_{\,r}^{m}$ and $\Phi_{\,r}^{m}$ \ represent
the first two (rightmost) curves$.$ Decoding is performed on a binary channel,
without using any soft-decision information.

These simulation results show that $\Phi_{\,r}^{m}$ gains about
$1$ dB over $\Psi_{\,r}^{m}$ \ on the code $\{_{2}^{8}\}$ and
about \ 0.5 dB on the code $\{_{3}^{8}\}$ even for high WER.  A
subsequent improvement can be obtained if we consider
\textit{soft-decision} decoding$,$ which recursively recalculates
the  \textit{posterior probabilities} of\ the new variables
obtained in both Steps 1 and 2. These  modifications of algorithms
$\Psi_{\,r}^{m}$ and $\Phi_{\,r}^{m}$ - called below
$\tilde{\Psi}_{\,r}^{m}$ and $\tilde{\Phi }_{\,r}^{m}$ - are
designed along these lines in \cite{dum4}. The simulation results
for the algorithm $\tilde{\Phi}_{\,r}^{m}$ are also presented in
Fig.~\ref{fig2} to Fig. \ref{fig4}, where these results are given
by the third curve.

This extra gain can be further increased if a few most plausible code
candidates are recursively retrieved and updated in all intermediate steps. We
note that the list decoding algorithms have been of substantial interest not
only in the area of error control but also in the learning theory. For
\textit{long} \textit{biorthogonal codes} $\left\{
\genfrac{}{}{0pt}{}{m}{1}%
\right\}  ,$ the pioneering randomized algorithm is presented in \cite{levi}.
For $m\rightarrow\infty$ and any constants $\varepsilon>0,$ $s>0,$ this
algorithm outputs a complete list of codewords located within the distance
$n(1-\varepsilon)/2$\ from any received vector, while taking only a polynomial
time \textsf{poly}$(ms/\varepsilon)$ to complete this task with high
probability $1-\exp\{-s\}.$ Substantial further advances are obtained for some
low-rate $q$-ary RM codes in \cite{gold} and the papers cited therein.

For binary RM codes of \textit{any order} $r,$ we mention three
different soft decision list decoding techniques, all of which
reduce the output WER at the expense of higher complexity. The
algorithm of \cite{sha} and \cite{sha1} reevaluates the most
probable information subblocks on a \textit{single run}. For each\
path $\xi,$ the decoder - called below $\tilde{\Psi}_{\,r}^{m}(L)$
- updates the list of $L$ most probable information subblocks
$\hat{\mathbf{a}}(\gamma)$ obtained on the previous paths
$\gamma.$ This algorithm has overall complexity of order $Ln\log
n.$  The technique of \cite{bos1} proceeds recursively at any
intermediate node,  by choosing $L$ codewords closest to the input
vector processed at this node. These lists are updated in
\textit{multiple recursive runs.} Finally, the third novel
technique \cite{sor1} executes sequential decoding using the main
stack, but also utilizes the complementary stack in this process.
The idea here is to lower-bound the minimum distance between the
received vector and the closest codewords that will be obtained in
the \textit{future steps}.

Computer simulations show that the algorithm of \cite{sha1}
achieves the best complexity-performance trade-off known to date
for RM codes of moderate lengths 128 to 512. In Fig.~\ref{fig2} to
Fig. \ref{fig4}, this algorithm $\tilde{\Psi}_{\,r}^{m}(L)$ is
represented by
the fourth curve, which shows a gain of about 2 dB over $\ \tilde{\Phi}%
_{\,r}^{m}$. Here we take $L=16$ in Fig.~\ref{fig2} and $L=64$ in
Fig.~\ref{fig3} and \ref{fig4}. Finally, complexity estimates
(given by the overall number of floating point operations) are
presented for all three codes in Table 1.

\begin{center}
\begin{tabular}
[c]{|c|c|c|c|c|}\hline Code & $\left|  \Psi_{r}^{m}\right|  $ &
$\left|  \Phi_{r}^{m}\right|  $ &
$\underset{}{\overset{}{|\tilde{\Phi}_{r}^{m}|}}$ & $|\tilde{\Psi}_{r}%
^{m}(L)|$\\\hline
$\overset{}{\underset{}{%
\genfrac{\{}{\}}{0pt}{}{7}{2}%
}}$ & 857 & 1264 & 6778 & 29602, $\,L=16$\\\hline
$\overset{}{\underset{}{%
\genfrac{\{}{\}}{0pt}{}{8}{2}%
}}$ & 1753 & 2800 & 16052 & 220285, $L=64$\\\hline
$\overset{}{\underset{}{%
\genfrac{\{}{\}}{0pt}{}{8}{3}%
}}$ & 2313 & 2944 & 12874 & 351657, $L=64$\\\hline
\end{tabular}
\end{center}\smallskip
Table 1. Complexity estimates for hard-decision algorithms
$\Psi_{r}^{m}$ and $\Phi_{r}^{m}$, and soft-decision versions
$\tilde{\Phi}_{r}^{m}$ and $\tilde{\Psi}_{r}^{m}(L)$.\medskip

Recall also that different information bits - even those retrieved
in consecutive steps - become much better protected as recursion
progresses. This allows one to improve code performance by
considering a subcode of the original code, obtained after a few
least protected information bits are removed. The corresponding
simulation results can be found in \cite{sha} and \cite{sha1}.

Summarizing this discussion, we outline a few important open problems.
\ Recall that the above algorithms $\Psi_{\,r}^{m}$ and $\Phi_{\,r}^{m}$ \ use
two simple recalculation rules%
\begin{equation}
\mathbf{y}\Longrightarrow\{\mathbf{y}^{\prime}\mathbf{y}^{\prime\prime
},(\mathbf{y}^{\prime}+\mathbf{y}^{\prime\prime})/2\}.\label{rule1}%
\end{equation}
Therefore the first important issue is to define whether any asymptotic gain
can be obtained:

- by changing the recalculation rules (\ref{rule1}) ;

- by using intermediate lists of small size $L;$

- by removing a few weakest information paths (bits).\smallskip

The second important problem is to obtain tight bounds on the decoding error
probability in addition to the decoding threshold derived above. This is an
open problem even for the simplest recalculations (\ref{rule1}) utilized in
this paper, let alone other rules, such as (\ref{boss}) or those outlined in
\cite{dum4}.

From the practical perspective, recursive algorithms show substantial promise
at the moderate lengths up to 256, on which they efficiently operate at
signal-to-noise ratios below 3 dB. It is also interesting to extend these
algorithms for low-rate subcodes of \ RM codes, such as the duals of the BCH
codes and other sequences with good auto-correlation.

In summary, the main result of the paper is a new probabilistic technique that
allows one to derive exact asymptotic thresholds of recursive algorithms.
Firstly, we disintegrate decoding process into a sequence of recursive steps.
Secondly, these dependent steps are estimated by independent events, which
occur when all preceding decodings are correct. Lastly, we develop a
second-order analysis that defines a few weakest paths over the whole sequence
of consecutive steps.  \medskip

\textit{Acknowledgement. }The author thanks K. Shabunov for helpful
discussions and assistance in computer simulation.

\begin{onecolumn}
\setcounter{figure}{2}
\begin{figure}
\begin{center}
\includegraphics[width=4.5in]{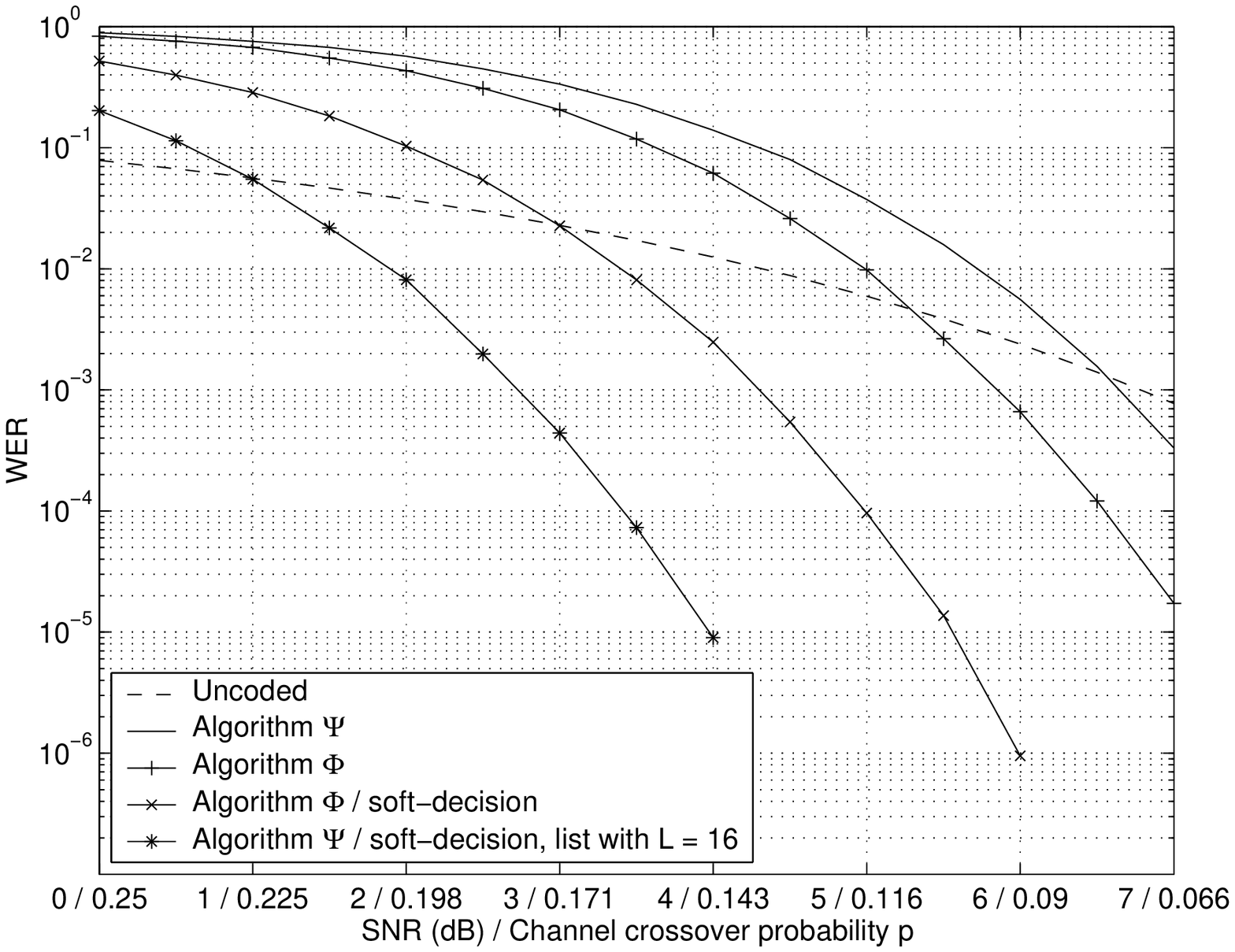}
\end{center}
\caption{ \normalsize {$\left\{ \genfrac{}{}{0pt}{}{7}{2}\right\}
$ RM code, $n=128$, $k=29$. \ \
Code word error rates (WER) for hard-decision algorithms
$\Psi_{\,r}^{m}$ and $\Phi_{\,r}^{m}$, and soft-decision
algorithms $\tilde{\Phi}_{\,r}^{m}$ and
$\tilde{\Psi}_{\,r}^{m}(L)$ (list of size $L=16.)$ }} \label{fig2}
\end{figure}
\begin{figure}
\begin{center}
\includegraphics[width=4.5in]{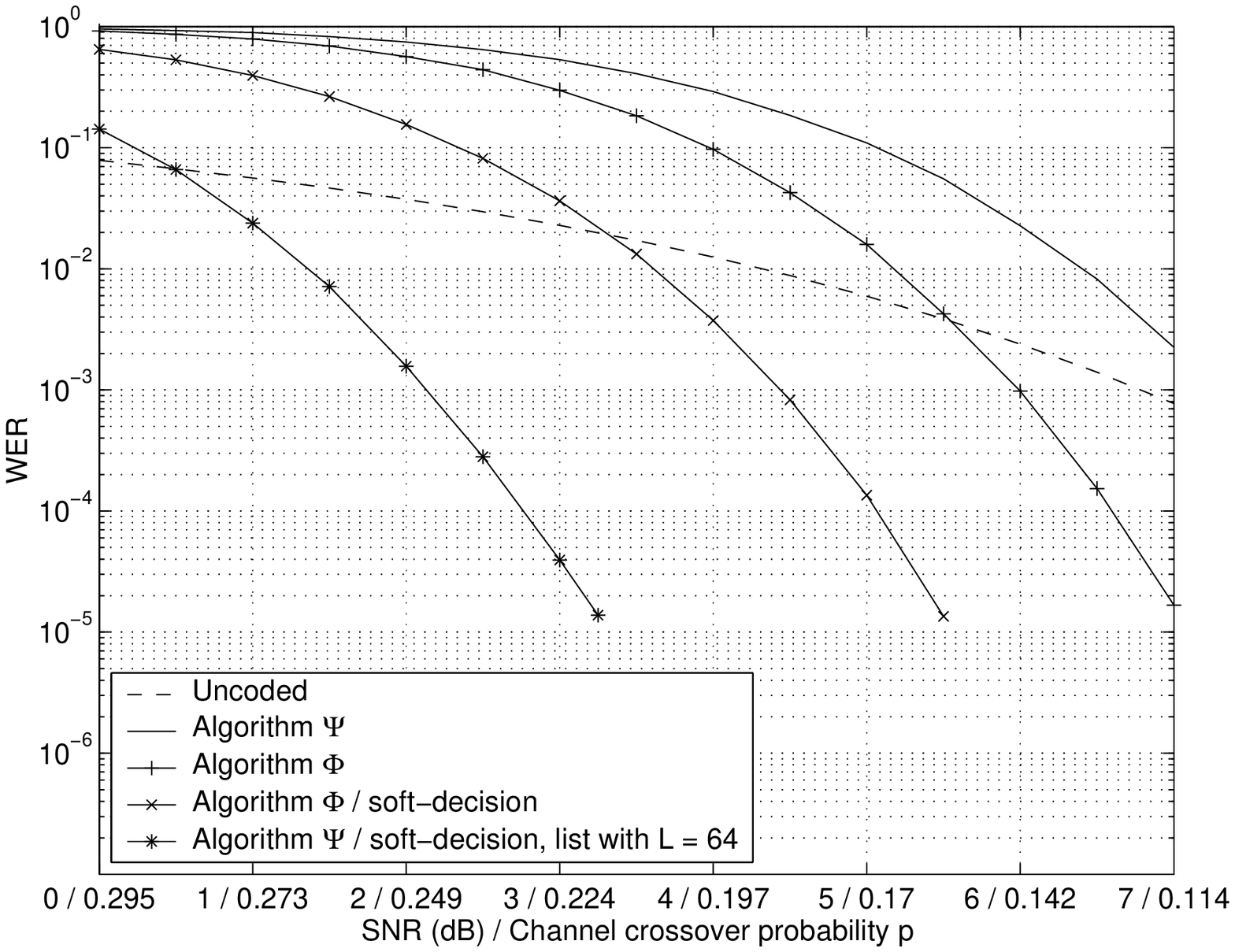}
\end{center}
\caption{ \normalsize { $\left\{ \genfrac{}{}{0pt}{}{8}{2}\right\}
$ RM code, $n=256$, $k=37$.\ \
Code
word error rates (WER) for hard-decision algorithms
$\Psi_{\,r}^{m}$ and $\Phi_{\,r}^{m}$, and soft-decision
algorithms $\tilde{\Phi}_{\,r}^{m}$ and
$\tilde{\Psi}_{\,r}^{m}(L)$ (list of size $L=64.)$ }} \label{fig3}
\end{figure}
\begin{figure}
\begin{center}
\includegraphics[width=4.5in]{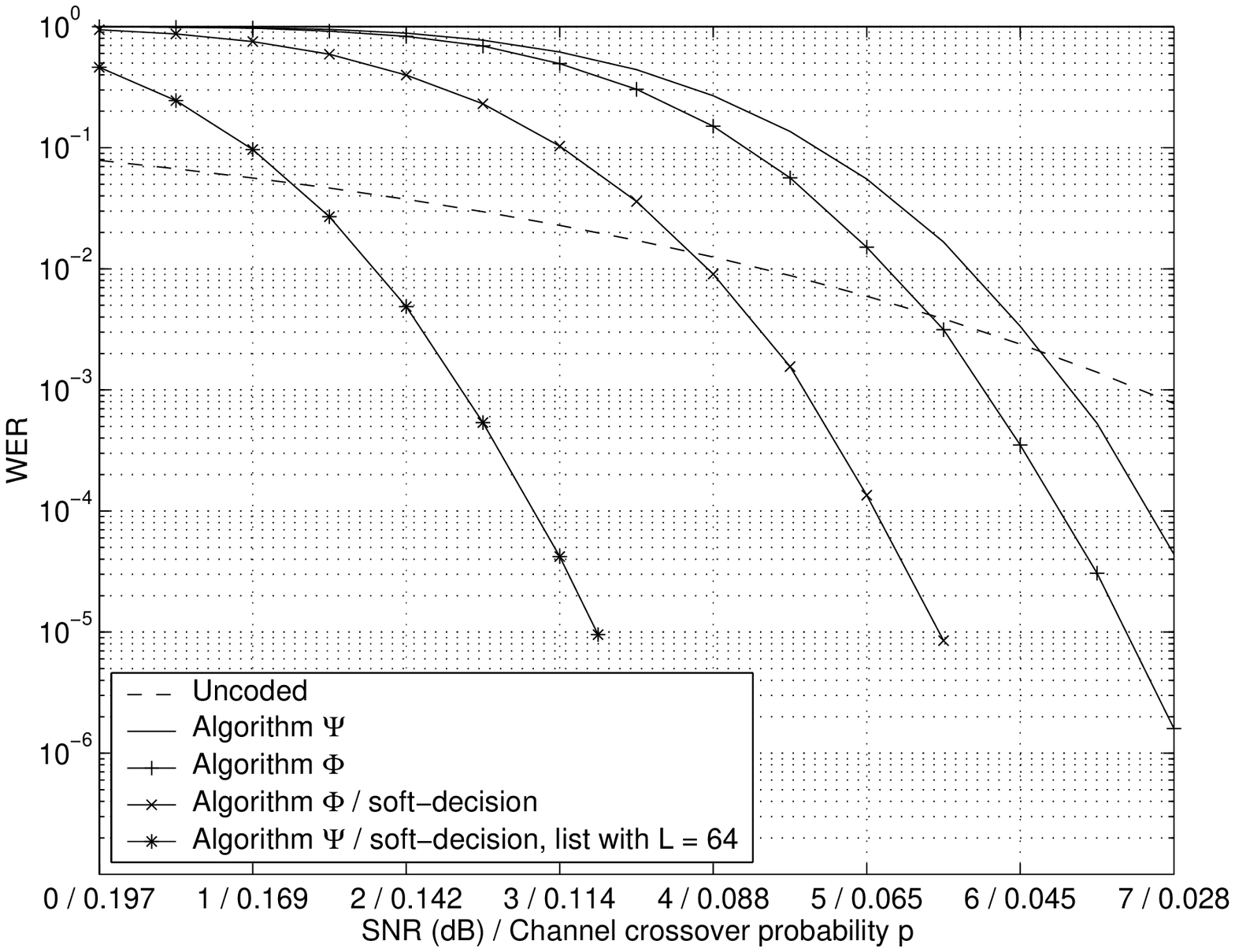}
\end{center}
\caption{ \normalsize { $\left\{ \genfrac{}{}{0pt}{}{8}{3}\right\}
$ RM code, $n=256$, $k=93$. \ \
Code
word error rates (WER) for hard-decision algorithms
$\Psi_{\,r}^{m}$ and $\Phi_{\,r}^{m}$, and soft-decision
algorithms $\tilde{\Phi}_{\,r}^{m}$ and
$\tilde{\Psi}_{\,r}^{m}(L)$ (list of size $L=64.)$ }} \label{fig4}
\end{figure}
\end{onecolumn}
\end{document}